\documentclass[aps,pra,twocolumn,superscriptaddress,10pt,article,nofootinbib,showpacs,longbibliography]{revtex4-1}
\usepackage{amsfonts}
\usepackage{amsmath}
\usepackage{amsthm}
\usepackage{braket}
\usepackage{graphicx}
\usepackage{hyperref}
\usepackage[dvipsnames]{xcolor}
\usepackage{amssymb}
\usepackage{dsfont}
\usepackage{verbatim}
\usepackage{amsmath,amscd}
\usepackage{multirow}
\usepackage[normalem]{ulem}
\usepackage{bm}
\usepackage[font=footnotesize]{caption}
\usepackage{subcaption}
\def \k{{\mathbf k}}
\def \q{{\mathbf q}}
\def \Q{{\mathbf Q}}
\def \A{{\mathbf A}}
\def \r{{\mathbf r}}
\def \R{{\mathbf R}}
\def \T{{\mathcal T}}
\def \beq{\begin{eqnarray}}
\def \eeq{\end{eqnarray}}
\def \nn{\nonumber \\}
\DeclareMathOperator{\Tr}{Tr}

\begin{document}

\title{A mechanism for anomalous Hall ferromagnetism in twisted bilayer graphene}

\author{Nick Bultinck}
\thanks{N.B. and S.C. contributed equally to this work.}
\affiliation{Department of Physics, University of California, Berkeley, CA 94720, USA}
\author{Shubhayu Chatterjee}
\thanks{N.B. and S.C. contributed equally to this work.}
\affiliation{Department of Physics, University of California, Berkeley, CA 94720, USA}
\author{Michael P. Zaletel}
\affiliation{Department of Physics, University of California, Berkeley, CA 94720, USA}
\affiliation{Materials Sciences Division, Lawrence Berkeley National Laboratory, Berkeley, California 94720}

\begin{abstract}
Motivated by the recent observation of an anomalous Hall effect in twisted bilayer graphene, we use a lowest Landau level model to understand the origin of the underlying symmetry-broken correlated state. This effective model is rooted in the occurrence of Chern bands which arise due to the coupling between the graphene device and its encapsulating substrate. Our model exhibits a phase transition from a spin-valley polarized insulator to a partial or fully valley unpolarized metal as the bandwidth is increased relative to the interaction strength, consistent with experimental observations. In sharp contrast to standard quantum Hall ferromagnetism, the Chern number structure of the flat bands precludes an instability to an inter-valley coherent phase, but allows for an excitonic vortex lattice at large interaction anisotropy.
\end{abstract}

\maketitle

Moir\'e graphene systems are a class of simple van der Waals heterostructures \cite{hetero} hosting interaction driven low-energy physics, making them an exciting platform to advance our understanding of correlated quantum matter. In twisted bilayer graphene (TBG) with a small twist angle between adjacent layers, interaction effects are enhanced by van Hove singularities coming from 8 nearly flat bands around charge neutrality (CN) in the Moir\'e- or mini-Brillouin zone (mBZ) 
%with a very small bandwidth
\cite{Bistritzer,Mele,Lopes,CastroNeto,Trambly,Shallcross,Morell,Moon,Li,Luican,Yan,Brihuega,Ohta,Havener,Morell,Gail,Uchida,Sboychakov,Jung2,Wong,Fang}. Observation of correlated insulating states when 2 or 6 of the 8 TBG flat bands are filled confirms the importance of interactions \cite{Cao3,Kim,Cao,Yankowitz,Kerelsky,Choi,Chen}. 
%A confirmation of the important role played by interactions in these mBZ flat bands was provided in Ref.~\cite{Cao3} and Refs. \cite{Kim,Cao,Cao2,Yankowitz,Kerelsky,Choi}, where interaction-dominated gaps were observed when 2 or 6 (filling $\nu = -2, 2$) of the 8 flat bands in TBG are filled. 
%Interestingly, at densities near some of these Mott insulators the system becomes superconducting \cite{Cao2,Yankowitz}.

Recent experiments indicate that certain magic angle graphene devices have large resistance peaks at $\nu = 0, 3$, with the latter featuring an  anomalous Hall (AH) effect detected via hysteresis  in the Hall conductance as a function of the out-of-plane magnetic field \cite{Goldhaber}.
The Hall conductance is of order $e^2 / h$ but not yet quantized.
Some have detected an meV-scale gap at CN, and a hysteretic behaviour of the Hall conductance with applied field at $\nu = -1$ \cite{Efetov}.
In this work we discuss how the  breaking of the 180-degree rotational symmetry  ($C_{2z}$) by a partially aligned hexagonal boron-nitride (h-BN) substrate could explain these observations. 
A variety of works \cite{Jung,Hunt13,Amet13,Lee2016,1707.09054, 1712.01968, Kim2018} have found that h-BN opens up a band gap at the Dirac points of graphene whose magnitude depends on the graphene / h-BN alignment angle, reaching $\Delta_{AB} \sim 17$meV \cite{Kim2018} to $\sim 30$meV \cite{1707.09054, 1712.01968} at perfect alignment.
Notably, even in seemingly unaligned devices with little or no observable h-BN induced Moir\'e potential, band gaps of several meV are still observed \cite{Kim2018, 1712.01968}.
In TBG, the substrate can likewise gap out the band Dirac points at the $K_{\pm}$ points of the mBZ, splitting the bands as $8 = 4 + 4$ to create a gap at CN. We find that for certain sublattice splittings the resulting flat bands have Chern number $C = \pm 1$. This makes the TBG case similar to ABC stacked trilayer graphene, where under an appropriately directed electric field the flat bands have Chern numbers $\pm 3$ \cite{Senthil}.

\begin{figure}
\centering
\begin{subfigure}[t]{0.2\textwidth}
\includegraphics[width=\textwidth]{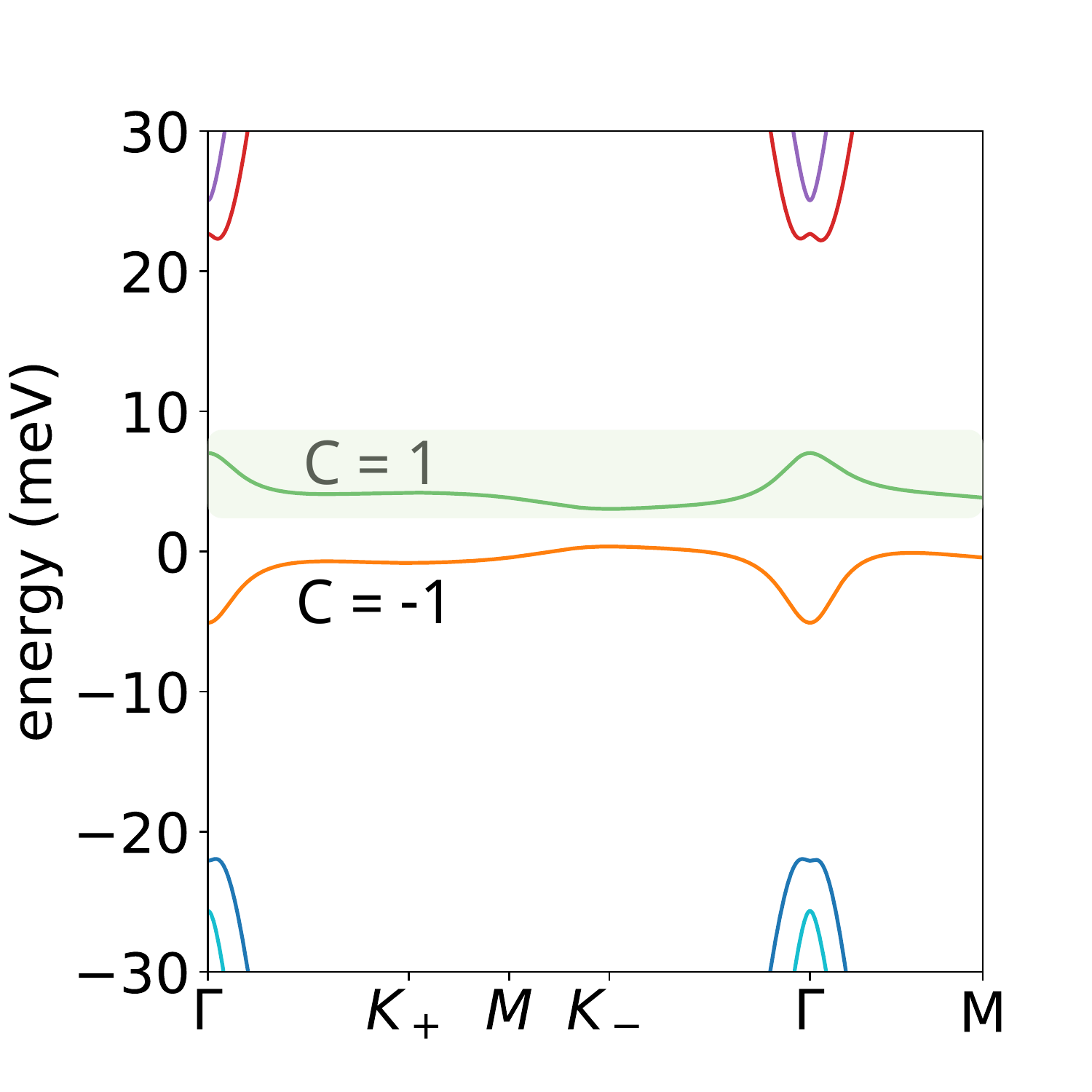}\caption{}
\end{subfigure}
\begin{subfigure}[t]{0.24\textwidth}
\includegraphics[width=\textwidth]{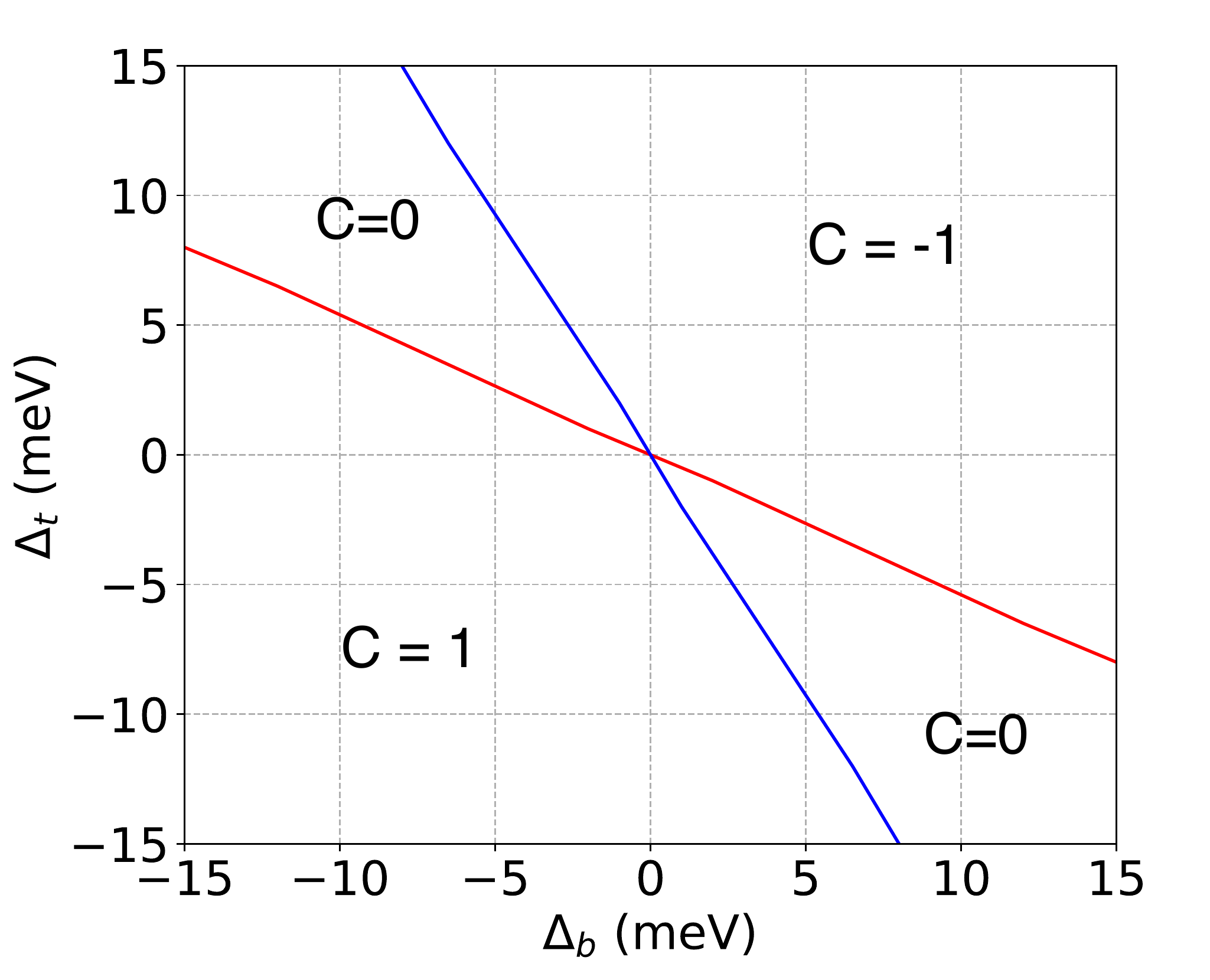}\caption{}
\end{subfigure}
\caption{The effect of sublattice splittings $\Delta_t$ and $\Delta_b$ on the spinless single-valley Moir\'e Hamiltonian (SVMH). (\textbf{a})  Band structure around CN for $\Delta_t=15$ meV and $\Delta_b=0$. The flat band above (below) CN has Chern number $C=-1$ ($C=1$). (\textbf{b}) Phase diagram of the SVMH for different $\Delta_t$ and $\Delta_b$. Phases are labeled by the Chern number $C$ of the flat $\tau=+$ conduction band. Blue (red) transition lines are characterized by a Dirac cone at the $K_-$ ($K_+$) point of the mBZ.} \label{fig:Chern}
\end{figure}

Accounting for the $C_{2z}$-breaking substrate, the basic structure of the problem is  as follows. The gap at CN allows us to focus only on the four nearly degenerate conduction (valence) bands for fillings above (below) CN, i.e, $\nu > 0$ ($\nu < 0$). These four Chern bands are uniquely labeled by their valley $\tau = +, -$ and spin $s = \uparrow, \downarrow$; time-reversal switches the valley index and enforces opposite Chern numbers for bands from opposite valleys. Since a $|C|=1$ band is topologically equivalent to a Landau level (LL), the problem is roughly analogous to a spinful bilayer quantum Hall problem with one flux quanta per unit cell, but with opposite layers (valleys) experiencing opposite magnetic fields. The LLs are degenerate, but as in a quantum Hall ferromagnet (QHFM)\cite{Sondhi} at integer filling the electrons may open a gap by spontaneously polarizing into a subset of these LLs, or a coherent superposition of them. 
In conventional quantum Hall bilayers at filling $\nu=1$, interactions generically drive inter-layer coherence, e.g., the exciton condensate \cite{Eisenstein14,Tutuc04}. But the twist here is the opposing Chern numbers of the two valleys. We find that the Chern number structure provides a topological reason for penalizing a coherent state: an exciton condensate between $C=1, -1$ bands is analogous to a superconductor in a strong magnetic field, which forces vortices into the order parameter, reducing the gain in the correlation energy. Hence, a spontaneously valley-polarized (VP) state is stable and exhibits AH effect with Hall resistance $\sim h/e^2$ (QAH if completely spin and valley polarized). Further, pinning of valley-polarization by an out-of-plane $B^z$ due to a large orbital g-factor explains the  presence of the $R_{xy}$ hysteresis loop observed in Ref.~\cite{Goldhaber}.

The possibility of spin and valley polarization and/or quantum anomalous Hall physics and chiral edge states in TBG has been discussed previously in Refs. \cite{Senthil,Dodaro,Thomson,KangVafek,Ochi,Xie,Xu,Nandkishore, LiuDai,Senthil2}, albeit from a different perspective.
%In particular, Ref.~\cite{Senthil} studied stacks of multilayer graphene twisted relative to each other (eg. bilayer on bilayer), or multilayer graphene with a Moir\'e potential from a h-BN substrate. They discuss several possible phases for nearly flat bands at integer fillings, and our findings of a stable QAH state at filling $\nu = 3$ are in agreement.
We also note that a recent self-consistent Hartree-Fock (HF) treatment of the continuum model exhibits \emph{spontaneous} $C_{2z}T$ breaking at CN, though the resulting Chern numbers were $C = \pm 2$ \cite{Xie}. %the authors also point out the possibility of QAH effects at odd fillings ($\nu = \pm 1, \pm 3$).  

\emph{Substrate-induced Dirac mass and Chern numbers--} We model the effect of the h-BN substrate \cite{Jung} by including in our band calculations a uniform but $C_{2z}$ breaking A-B sublattice splitting $\Delta_t$ and $\Delta_b$ on the top and bottom layer respectively (see \cite{supplementary} for details). While h-BN may also introduce a Moir\'e potential, its magnitude falls off much more rapidly with alignment angle than $\Delta_{t/b}$ \cite{Kim2018}.
For our calculations we used a twist angle $\theta\approx 1.05^\circ$, and have taken a phenomenological corrugation effect into account by using a larger AB/BA inter-layer hopping $w_1$ as compared to the AA/BB inter-layer hopping $w_0$. Taking $w_0/w_1=0.85$ results in flat bands separated from the dispersing bands by an energy gap of approximately $20$ meV (for zero sublattice splittings). 

With sublattice splitting, the phases of the $\tau = +$ valley (or $K$-valley of monolayer graphene) Moir\'e Hamiltonian for different parameter regimes of $\Delta_t$ and $\Delta_b$ are shown in Fig.~\ref{fig:Chern}. We find four different regions where both Dirac cones in the mBZ are gapped because of the sublattice splittings. In these regions, there are two isolated flat bands. We find that these four regions have bands with Chern numbers \cite{Hatsugai}
$C=\pm 1$ or $C=0$, and are separated from each other by a Dirac point at either the $K_-$ or $K_+$ point in the mBZ. In Fig.~\ref{fig:Chern} we show the Chern number of the flat band for the $\tau = +$ valley above (below) CN in green (orange). The Chern number for the flat bands from the $\tau = -$ valley can be obtained by time-reversal.

The location of the $C=\pm1$ phases can be understood from the fact that for small $\Delta_t=\Delta_b>0$ or $\Delta_t=\Delta_b<0$, the leading order effect of the sublattice potentials is to generate Dirac masses with the same sign at both the $K_-$ and $K_+$ points of the mBZ. Because both Dirac cones in a single valley have the same chirality, this leads to bands with Chern number $\pm 1$, a feature earlier work dubbed a ``flipped Haldane model''\cite{Zou} (see also \cite{Po,Po2,Bernevig}). From Fig.~\ref{fig:Chern} we see that even if only one of the layers has a non-zero sublattice splitting, the strong inter-layer coupling ensures that both Dirac cones at the mBZ $K$-points acquire a mass.%, and that the bands have non-zero Chern number.
These findings can also be inferred analytically within the ``chiral'' approximation of tBLG \cite{SanJose,Tarnopolsky}, in which all bands are sub-lattice polarized and carry Chern number $C = \sigma \tau$, where $\sigma$ denotes sublattice.

\emph{Metal - valley polarization competition--} In this work, we focus only on the four flat \textit{conduction} bands above the CNP (the highlighted band in Fig.~\ref{fig:Chern} and its valley and spin counterparts). In the supplement, we numerically justify this for TBG, showing that $\Delta_t \sim 15$ meV ($\Delta_b = 0$) creates a $30$ meV gap between valence and conduction bands \cite{supplementary}. To phenomenologically model the effect of interactions in this set of bands we adopt a lowest Landau level (LLL) description. We can map the Chern bands to a LLL by constructing the Wannier-Qi states \cite{VanderbiltRMP2012,Qi,supplementary}. In the following, we use an approximation where the Wannier-Qi states of the flat bands are replaced by the continuum LLL wave functions of a two-dimensional electron gas. Physically, this amounts to neglecting the inhomogeneous Berry curvature in the Chern bands. The AH effect and  edge transport reported in Ref.~\onlinecite{Goldhaber} can be explained if there is one VP hole per Moir\'e unit cell. From the data in Ref. \cite{Goldhaber} is not possible to exclude a spin-unpolarized, gapless phase. If the spins do polarize however, the underlying mechanism is expected to be the same as in conventional QHFM \cite{Sondhi}, and is not sensitive to the opposite Chern numbers of the two valleys. Therefore, in the analysis below we ignore spin and focus on the mechanism of valley polarization. Considering the uniform repulsive nature of the projected Coulomb interaction and the numerical evidence against stripes in the LLL \cite{ND2004}, we disregard the possibility of interaction-induced charge density waves, and focus on the competition between valley-polarized, inter-valley coherent and metallic phases. For this we need to introduce two parameters in our LLL toy model: the bandwidth and the interaction anisotropy. To achieve a non-zero bandwidth we use a square lattice potential, that sidesteps the complexities of a hexagonal lattice and allows analytical progress.

We consider a torus of length $L_x$ ($L_y$) in the $x$ ($y$) direction, with a magnetic field perpendicular to the surface. We choose units in which $L_xL_y =2\pi N_\phi l_B^2 \equiv N_\phi a^2$, where $N_\phi$ is the number of flux quanta piercing the torus, and $l_B = (\hbar/e B)^{-1/2}$ is the magnetic length. In particular, we will take $L_x =  N_x a$ and $L_y =  N_y a$, with $N_\phi=N_xN_y$. Next to the magnetic field, we also add a periodic potential $V_P(x,y) = w(\cos(2\pi x/a) + \cos(2\pi y/a))$, such that there is exactly $2\pi$ flux in each unit cell. The potential is invariant under translations over $a$ in both the $x$ and $y$-direction, which means that the momenta $k_x=n\frac{2\pi}{N_x a}$ and $k_y=n \frac{2\pi}{N_y a}$ ($n\in\mathbb{Z}$) are good quantum numbers.

We are interested in the physics in the LLL with Chern numbers $C=1,-1$. The electron creation operator projected in these subspaces takes the form $\psi^\dagger_{\pm}(x,y)=\frac{1}{\sqrt{L_y l_B \sqrt{\pi}}} \sum_k e^{iky -\frac{1}{2l_B^2}(x\mp kl_B^2)^2}c_{\pm,k}^\dagger$, where we have chosen the Landau gauge which explicitly preserves (continuous) translation symmetry in the $y$-direction, such that $k= 2\pi n/L_y=2\pi n/N_y a$ with $n \in \{0,1,\dots, N_xN_y\}$. We now proceed in analogy to Ref.~\onlinecite{Mishmash}, and define the Bloch states $c^\dagger_{\pm,(k_x,k_y)}=c^\dagger_{\pm,\textbf{k}}$ as
\begin{equation}\label{bloch}
c_{\pm,\textbf{k}}^\dagger = \frac{1}{\sqrt{N_x}} \sum_{n=0}^{N_x-1} e^{\pm ik_x(k_y+nQ)l_B^2}c^\dagger_{\pm,k_y+nQ}\, ,
\end{equation}
where $Q=\sqrt{2\pi}/l_B = 2\pi/a$. The density operator in the LLL $n_{\pm}(\textbf{q})=\int\mathrm{d}\textbf{r} \, e^{-i\textbf{q}\cdot\textbf{r}} \psi^\dagger_{\pm}(\textbf{r})\psi_{\pm}(\textbf{r})$ takes the form
\begin{eqnarray}
n_{\pm}(\textbf{q}) & = & F(\textbf{q})\sum_{k_x,k_y} e^{\pm iq_y k_x l_B^2} c^\dagger_{\pm, \textbf{k}-\textbf{q}/2} c_{\pm,\textbf{k}+\textbf{q}/2}\, ,
\end{eqnarray}
where the form factor is given by $F(\textbf{q})=e^{-\textbf{q}^2l_B^2/4}$. In the Bloch basis, the Hamiltonian term associated with the periodic potential takes the diagonal form $H^p=\sum_{\textbf{k}} \varepsilon_\textbf{k} (c^\dagger_{+,\textbf{k}}c_{+,\textbf{k}} + c^\dagger_{-,\textbf{k}}c_{-,\textbf{k}})$, with $\varepsilon_{\textbf{k}} = -we^{-\pi/2}[\cos(k_x a) + \cos(k_y a)]$. 

We are interested in the effect of density-density interactions on the LLL electrons moving in the periodic potential, described by the following Hamiltonian: 
%Specifically, we add the following interaction term to the Hamiltonian:
\begin{eqnarray}
H^i & = & \frac{1}{2 N_\phi} \sum_{\textbf{q},\tau,\tau'}V_{\tau,\tau'}(\textbf{q}):n_\tau(\textbf{q}) n_{\tau'}(-\textbf{q}),
\end{eqnarray}
where we neglect the small inter-valley scattering terms \cite{supplementary}. We will consider a general repulsive interaction of the form $V(\q)F^2(\q)=u_0(\q)(\mathds{1}+\tau^x)+u_1(\q)(\mathds{1}-\tau^x)$. In analogy to quantum Hall ferromagnetism \cite{Sondhi,Eisenstein14,Ezawa2} and related strongly coupled systems \cite{refA1,refA2}, at half-filling of the two bands we expect that the main effect of $H^i$ is to introduce a valley Hund's coupling between the electrons resulting in an insulating ground state. On the other hand, the \textit{kinetic} term $H^p$ coming from the periodic potential favors a metal over the VP insulator. To study the competition between these two phases, we perform a HF analysis using Slater determinants with correlation matrix $\langle c^\dagger_{\tau,\textbf{k}}c_{\tau',\textbf{k}'}\rangle = \delta_{\tau,\tau'}\delta_{\textbf{k},\textbf{k}'}\Theta(\epsilon^\tau_F-\epsilon_{\textbf{k}})$, such that $\sum_\tau \sum_{\textbf{k}}\Theta(\epsilon^\tau_F-\epsilon_{\textbf{k}})=N_\phi$. The possibility of inter-valley coherent states is addressed in the next section. For each Slater determinant, we define the corresponding valley polarization $P_v$ as $P_v = (N_+-N_-)/N_\phi$, where $N_+$ ($N_-$) is the number of electrons in the $+$ ($-$) valley. Without loss of generality, we restrict to $P_v>0$.

%We first treat the case where the anisotropic part $u_1(\textbf{q})$ of the interaction is zero, and take for the
We first consider an isotropic ($u_1(\textbf{q}) = 0$) dual-gate screened Coulomb potential with LLL form factors $u_0(\textbf{q})=2\pi Ue^{- \textbf{q}^2l_B^2/2}\tanh{(d|\textbf{q}|)}/|\textbf{q}|$, and screening length $d=a$. % taken to be one Moir\'e lattice constant.  
Using this interaction potential, we calculated the HF energy $E^{HF}$ \cite{supplementary}. We find that for $W/U \lesssim 0.6$, where $W \equiv 4we^{-\pi/2}$ is the bandwidth, the completely VP state indeed has the lowest energy. When $W/U\approx 0.6$, the valley polarization $P_v$ of the optimal Slater determinant jumps and starts decreasing continuously, indicating a first-order Mott transition from the VP insulator to an itinerant valley-ferromagnet. Around $W/U\approx 2.0$, $P_v$ continuously goes to zero and a conventional metallic phase sets in \cite{supplementary}.

\emph{Inter-valley coherence and exciton vortex lattice--} In bilayer QH ferromagnets, the insulating layer-polarized state is 
% known to be 
unstable to a uniform exciton condensate or inter-layer coherent state in presence of infinitesimal interaction anisotropy $u_1(\textbf{q}) > 0$ \cite{Eisenstein14}. The situation here is different as even with $u_1(\textbf{q})=0$, there is no SU$(2)$ valley symmetry because of the Chern number mismatch. The VP state therefore only breaks discrete symmetries, 
% such that 
indicating there will be no instability of this insulating state. Another, more physical, way to understand the absence of an exciton condensation instability is to use an analogy with type II superconductors. Because electrons in bands with an opposite Chern numbers effectively see opposite magnetic fields, an electron-hole condensate $\Delta(\r) = \langle c^\dagger_{+,\r} c_{-,\r} \rangle$ will behave like a charge 2e superconducting order parameter in a perpendicular magnetic field. However, in our scenario a Meissner-like effect, corresponding to uniform amplitude of the exciton order parameter, is ruled out from the outset. Rather, the magnetic field must leak through vortices in the exciton order parameter, leading to an excitonic vortex lattice phase. In this section, we show 
%by a HF analysis 
that both the VP insulator and the unpolarized metal are energetically favorable to the exciton vortex lattice, for sufficiently small interaction anisotropy $u_1(\textbf{q})$.

\begin{center}
\begin{figure}
\includegraphics[scale=0.27]{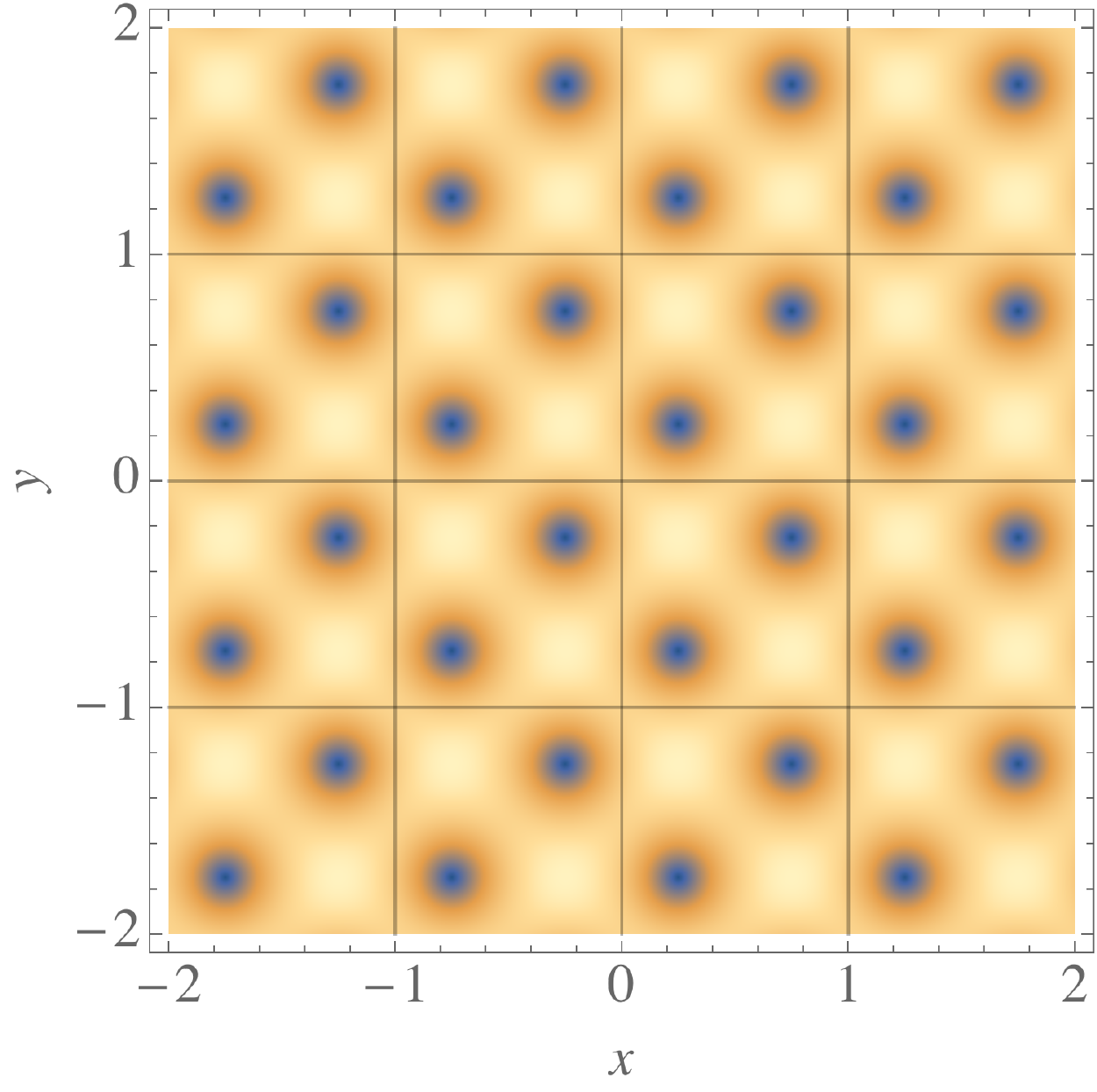}
\includegraphics[scale=0.29]{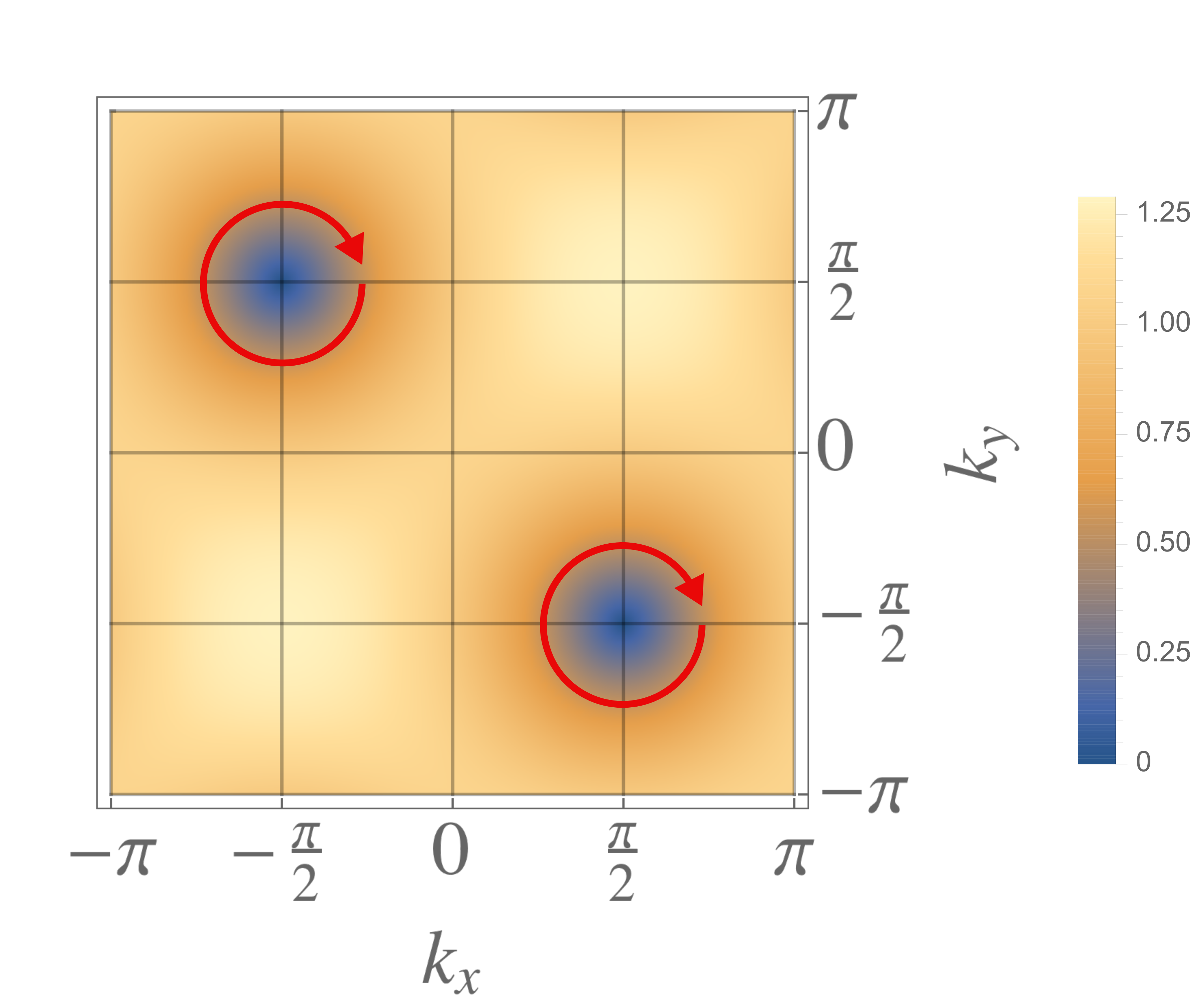}\caption{The magnitude of the excitonic order parameter in real (left) and momentum (right) space (for $a = 1$, $\Delta_0 = 1$). The red circles denote identical phase-winding of $\Delta_\k$ at both nodal points.}\label{fig:VL}
\end{figure}
\end{center}

For our LLL model, we can derive an exact expression for the exciton vortex lattice order parameter $\Delta(
\textbf{r})$.  To respect all symmetries of the square lattice, we expect $\Delta(\textbf{r})$ to have vortices at both the lattice sites and the plaquette centers, leading to a $4\pi$ vorticity in each unit cell. In the analytically tractable limit, we can uniquely determine $\Delta(\r)$ up to a translation by demanding its invariance under the magnetic translations $\T(a \hat{x})$ and $\T\left( \frac{a}{2}(\hat{x} + \hat{y}) \right)$, connecting the anticipated vortices \cite{supplementary}. In Fig. \ref{fig:VL} we plot the magnitude of $\Delta(\textbf{r})$ thus obtained, from which we clearly see the expected Abrikosov vortex lattice. Projecting $\Delta(\r)$ to the LLL Bloch basis wavefunctions $\phi_{\pm,\k}(\r)$ leads to a diagonal order parameter 
\beq
\Delta_\k = \Delta_0 \sum_{j = -\infty}^{\infty} e^{-i \frac{\pi}{2} j^2}  e^{- \frac{1}{4}(2 k_y + j Q)^2 l_B^2- i k_x (2 k_y + j Q )l_B^2}
\eeq
where $\Delta_0$ represents the overall strength of the exciton condensate. $\Delta_\k$ has two nodes with identical phase winding at $\k = \pm (\pi/2,-\pi/2)$, as shown in Fig. \ref{fig:VL} \cite{supplementary}.

The presence of two zeros in the BZ with the same phase winding is a topological \emph{requirement} for the exciton order parameter, and is not an artifact of our effective LLL model. In an isolated band $a$ with non-zero Chern number $C_a$, the phase of the electron creation  operator $c^\dagger_{a,\k}$ cannot be chosen to be both continuous and single-valued over the BZ. In particular, it must wind $2 \pi C_a$ times along the boundary of the BZ in a continuous gauge choice. This implies that the phase of $\Delta_\k = \langle c^{\dagger}_{+,\k} c_{-,\k} \rangle$ winds $2\pi(C_a - C_b) = 4\pi$ times along the BZ boundary for bands from opposite valleys with $C_a = 1$ and $C_b = -1$, which precisely corresponds to winding around two zeros with identical chirality.

\begin{figure}
\centering
\begin{subfigure}[t]{0.27\textwidth}
\includegraphics[width=\textwidth]{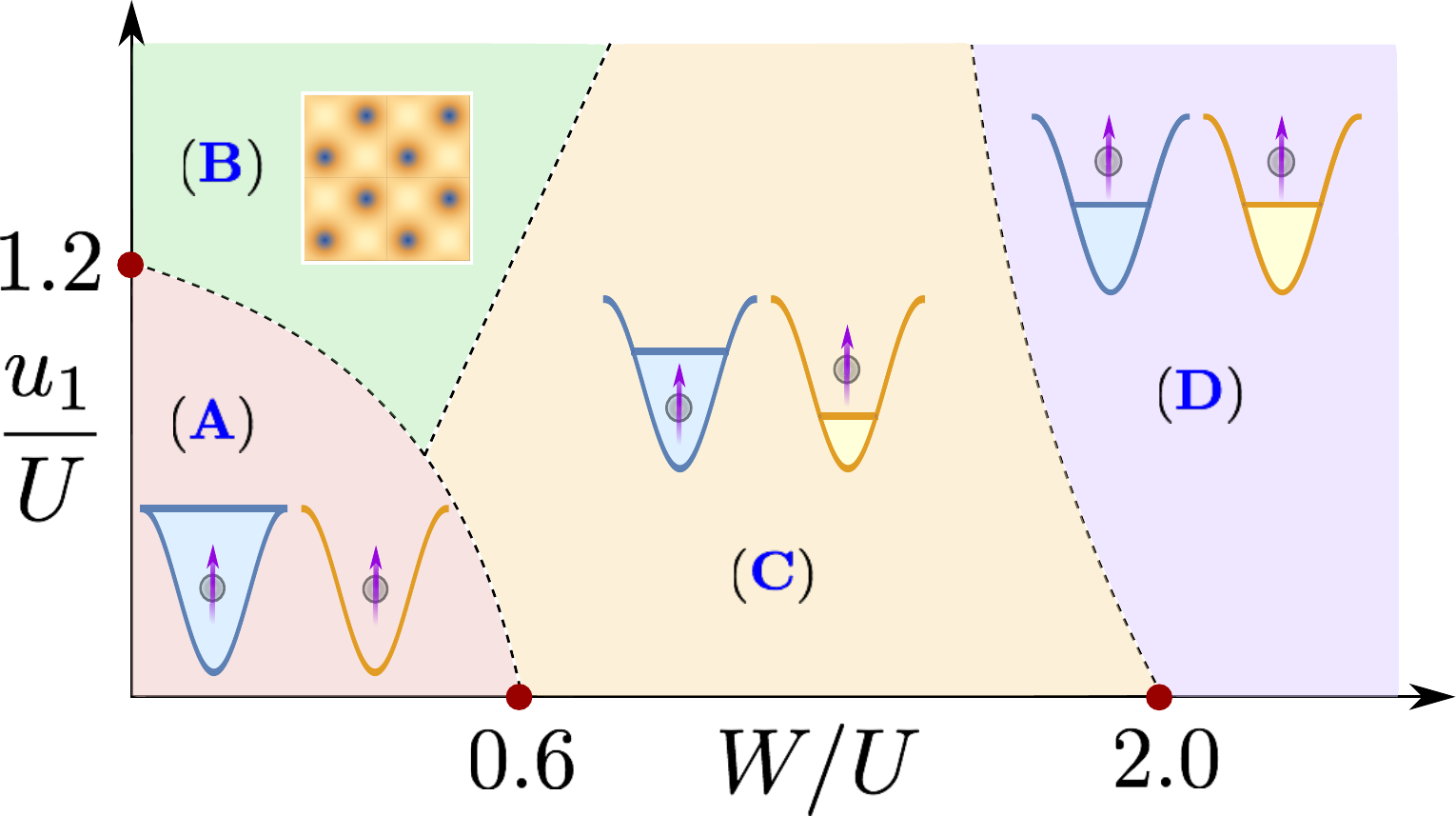}\caption{}
\end{subfigure}
\begin{subfigure}[t]{0.20\textwidth}
\includegraphics[width=\textwidth]{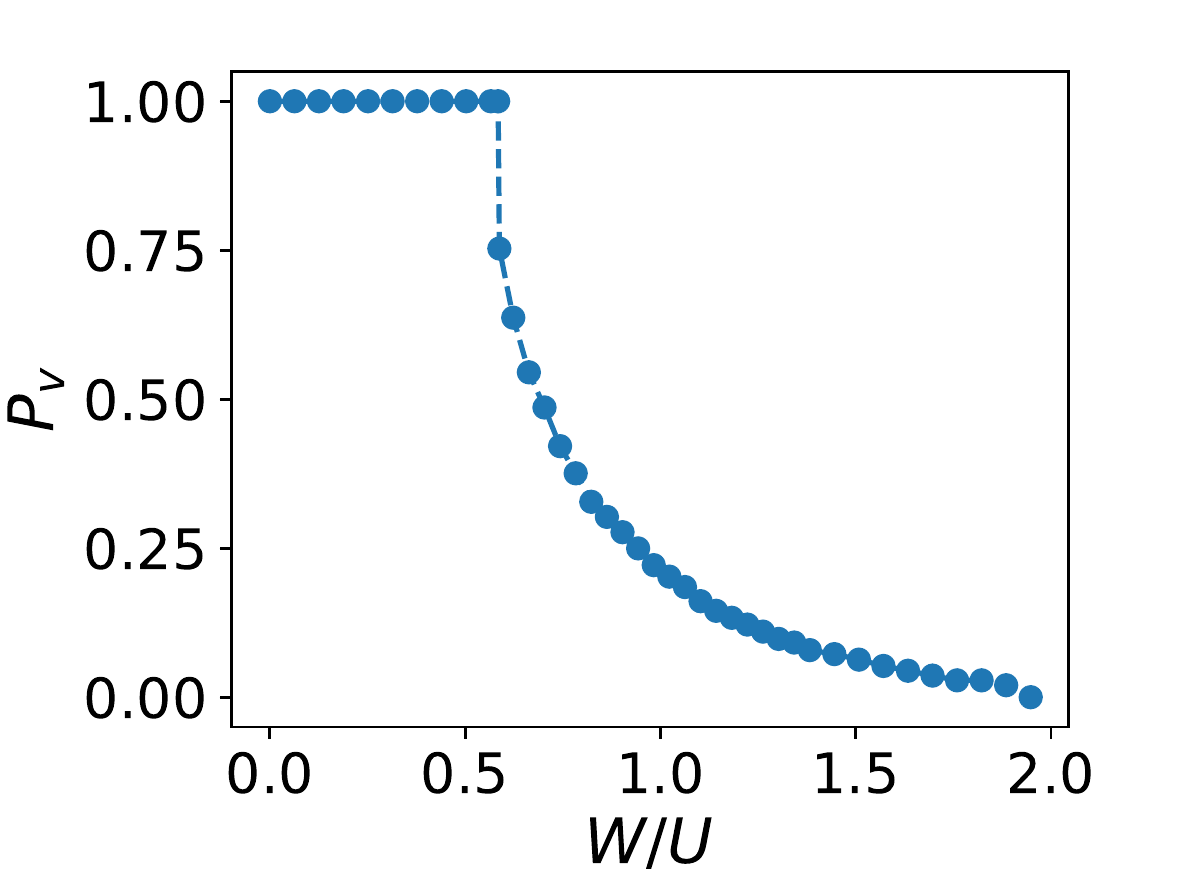}\caption{}
\end{subfigure}
\caption{(a) Approximate phase diagram of spin-polarized interacting electrons from opposite valleys in $C = \pm 1$ bands%, as a function of bandwidth $W$ and interaction anisotropy $u_1$ relative to the isotropic interaction $U$
. The phases are (A) fully VP insulator, (B) exciton vortex lattice, (C) partially polarized metal or itinerant valley-ferromagnet, and (D) unpolarized metal. Everywhere within phases A and C, %is expected to exhibit zero-field hysteresis in 
$R_{xy}\neq 0$. (b) Metal-insulator competition and the valley polarization $P_v$ %as a function of W/U 
for isotropic interaction.
} \label{fig:PD}
\end{figure}

We now demonstrate that variational states with an exciton vortex lattice have higher energy than the VP state or the metal for small anisotropy $u_1$ in the interaction $H^i$. We consider the Slater determinant ground state $\ket{\psi_{MF}}$ of the mean-field Hamiltonian $H_{MF} = \sum_{\k,\tau,\tau^\prime} c^\dagger_{\k,\tau} h_{\tau,\tau^\prime}(\k) c_{\k,\tau^\prime}$, where $h_{\tau,\tau'}(\textbf{k})=\epsilon_{\k}\mathds{1}+h\tau^z+\text{Re}(\Delta_{\textbf{k}})\tau^x+\text{Im}(\Delta_{\textbf{k}})\tau^y$. $\ket{\psi_{MF}}$ is characterized by the valley polarization $P_v$ (determined by $h$) and an exciton vortex lattice of strength $\Delta_0$, to be treated as variational parameters. The correlation matrix evaluated in this state takes the form of the projector $\langle c^\dagger_{\tau, \k} c_{\tau^\prime \k^\prime} \rangle = P_{\tau, \tau^\prime}(\k) \delta_{\k,\k^\prime}$, which can be used to evaluate the regularized HF energy density $e^{HF}(P_v,\Delta_0)$ of the variational state for a given microscopic interaction at a fixed filling $\nu = 1$. We find that the global minimum of $e^{HF}$ lies at $|P_v| = 1$ and $\Delta_0 = 0$ for the insulator in the limit of flat bands and isotropic interaction ($u_1 = 0$) \cite{supplementary}. We next show that the states of interest, with a fixed valley polarization $P_v$ at filling $\nu = 1$, are stable to the formation of an vortex lattice in presence of small interaction anisotropy. To do this, we consider the difference in energy density $e^{HF}(P_v,\Delta_0) - e^{HF}(P_v,0)$ perturbatively in $|\Delta_0|$ for arbitrary repulsive interaction parametrized by $u_0$ and $u_1$; a positive difference would indicate that $\Delta_0 = 0$ corresponds to an energy minimum. For the polarized phase, we find 
\beq
e^{HF}(1,\Delta_0) - e^{HF}(1,0) = \frac{1}{8 h^2} \bigg[ \int_{\k,\q} u_0(\q) |\Delta_{+} - \Delta_{-}|^2 \nn + \int_{\k,\q}  u_1(\q)|\Delta_{+} + \Delta_{-}|^2
 - 4 u_1(\mathbf{0}) \int_{\k} |\Delta_\k|^2  \bigg]\, ,~~~
\eeq
where $\Delta_{\pm} \equiv \Delta_{\k \pm \q/2}$ \cite{supplementary}. For a uniform exciton condensate, $\Delta_\k = \Delta_0$ and this energy difference is negative \cite{supplementary}. However, for an exciton order parameter formed with electrons and holes from opposite Chern bands, %the flux necessitates a vortex lattice and 
$\nabla_\k \Delta_\k \neq 0$. Therefore, when $u_1$ is sufficiently small compared to $u_0$ %(irrespective of their exact microscopic form) 
the energy of the state with non-zero $\Delta_\k$ is higher. %This indicates that 
So the VP state with $\Delta_0 = 0$, previously shown to be the ground state with an isotropic interaction for small $W/u_0$, is indeed robust to small interaction anisotropy. Analogous computations \cite{supplementary} show that the unpolarized metal ($P_v = 0 = \Delta_0$) is stable to the vortex lattice as well. An approximate phase diagram of our model for a short-range (LLL-projected) interaction anisotropy $u_1(\q) = u_1 e^{- \textbf{q}^2l_B^2/2}$ is presented in Fig.~\ref{fig:PD}. For TBG, we expect $W/U \lesssim 0.2$ from the ratio of the bandwidth to the Coulomb interaction, and the anisotropy $u_1/U \lesssim 0.01$ to be small \cite{supplementary,CBZ2019}, indicating a VP phase consistent with experiments \cite{Goldhaber,Serlin2019}. In the supplement, we numerically solve the mean-field equations for TBG on hBN at $\nu=3$ and confirm that the spin and VP QAH state is indeed the ground state.

%\footnote{The only assumption in the numerics is that translation symmetry is not broken}.

\emph{Valley Zeeman effect--} Having argued in favor of a VP state at $\nu =3$, we turn to the observed hysteresis in the $\nu = 3$ Hall conductance as a function of out-of-plane magnetic field $B^z$ ~\cite{Goldhaber}. To this end, we compute the orbital $g_v$-factor for the TBG conduction bands. In a band $\tau$ without time-reversal electrons can carry a momentum-dependent orbital moment $m_{\tau,\textbf{k}} $ \cite{XiaoNiu,ChangNiu}.
%Here $\tau$ indexes valley, so  t
Time-reversal ensures that $ m_{\tau,\textbf{k}} = -m_{-\tau,-\textbf{k}}$, which averaged over the mBZ produces a valley-Zeeman splitting $E = -g_v \frac{\tau^z}{2} \mu_B B^z $.
%We find $g_v$ \cite{supplementary} and 
We find that for $\Delta_b = 0, \Delta_t \sim 10-30$ meV,  $g_v $ ranges from approximately -2 to -6 \cite{supplementary}. 
%A similar conclusion was recently obtained for ABC stacked trilayer graphene \cite{Senthil2}.
Note that for $B^z >0$, the $C=1$ band comes down in energy. The sign of this effect is in agreement with the Landau fans of Refs.~\cite{Goldhaber,Serlin2019}.

\emph{Conclusion--} We showed that broken inversion symmetry in TBG due to substrate (h-BN) coupling leads to two Chern bands per valley. Spontaneous polarization of holes in spin and valley space then leads to an AH state at $\nu = 3$. Using a LLL model, a HF analysis establishes a stable VP state as the ground state when the bandwidth is small compared to the interaction strength. The opposite Chern numbers for the two valleys precludes uniform inter-valley coherence. The resultant exciton vortex lattice structure reduces correlation energy gain and stabilizes valley-polarization. This result agrees with numerical work on a Hubbard model \cite{NeupertSantos}.

\emph{Note added--} Recently, a quantized AHE with net Chern number $C = 1$ has been observed for a gapped insulator at $\nu = 3$ in TBG aligned with h-BN \cite{Serlin2019}, consistent with our theoretical results. Quantized AHE arising from valley-Chern bands have also been observed \cite{Chen2019} and proposed \cite{Liu2019,Lee2019} in other Moir\'e heterostructures, in accordance with our phenomenological picture of interaction in nearly flat bands with opposite Chern numbers. 

\emph{Acknowledgements--} We thank Aaron Sharpe, Eli Fox and David Goldhaber-Gordon for discussions about their data and sharing their insights. We also thank Ryan Mishmash for explaining the details of Ref.~\onlinecite{Mishmash} to two of us (SC and NB), and Ehud Altman, Senthil Todadri and Andrea Young for inspiring discussions. Our work overlaps with concurrent work by Y. Zhang, D. Mao and T. Senthil 
\cite{Senthil3}. MZ and NB were supported by the DOE, office of Basic Energy Sciences under contract no. DE-AC02-05-CH11231. SC acknowledges support from the ERC synergy grant UQUAM via E. Altman.

\let\oldaddcontentsline\addcontentsline% Store \addcontentsline
\renewcommand{\addcontentsline}[3]{}% Make \addcontentsline a no-op
\bibliography{bibliography}
\let\addcontentsline\oldaddcontentsline

\pagebreak
\onecolumngrid
\begin{center}
\large{\textbf{SUPPLEMENTARY MATERIAL}}
\end{center}
\tableofcontents

\section{Flat bands with sublattice splitting}

As was shown in Ref. \cite{Jung,Hunt13,Amet13}, the h-BN substrate generates a substantial Dirac mass when it is nearly aligned with the graphene sheet. We model this by introducing a $C_{2z}$ symmetry breaking sublattice-staggered potential $\Delta_t$ and $\Delta_b$ in respectively the top and bottom layer graphene sheet. 

\begin{center}
\begin{figure}
a) 
\includegraphics[scale=0.5]{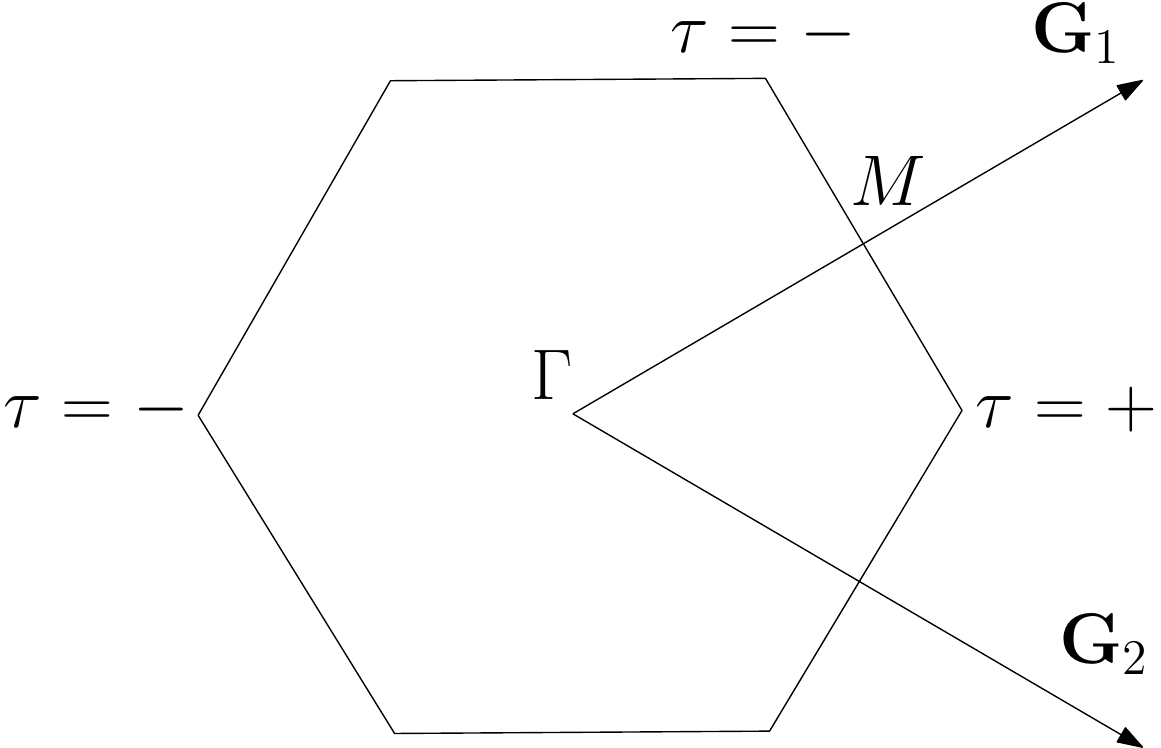} \hspace{1.5 cm} b)
\includegraphics[scale=0.6]{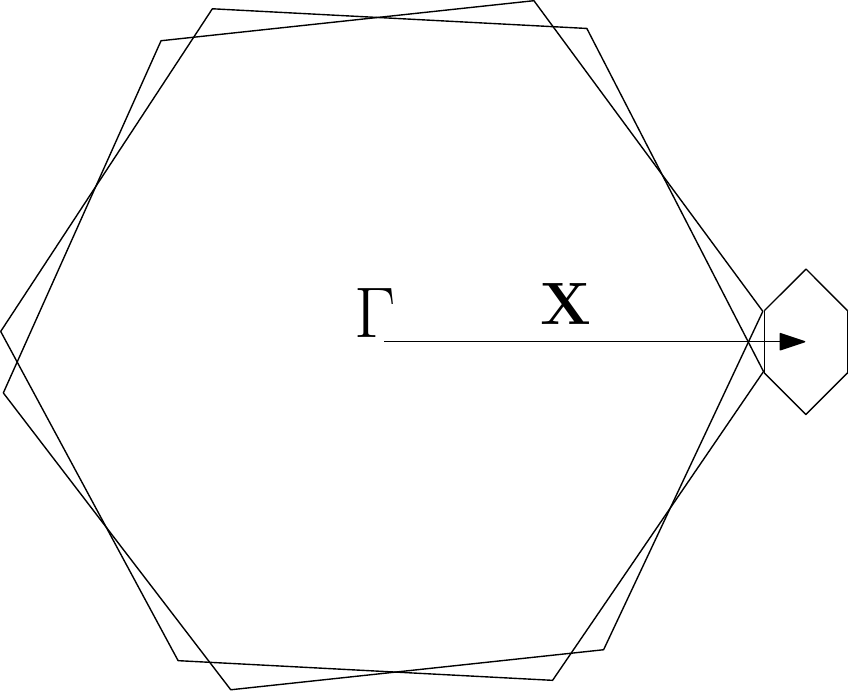}\caption{(a) The mono-layer graphene Brillouin zone with the two basis vectors $\textbf{G}_1$ and $\textbf{G}_2$ of the reciprocal lattice. We have indicated the high-symmetry $K_+$ and $K_-$ points, where the Dirac cones are located. (b) The mono-layer Brillouin zones of the top and bottom graphene layer with relative twist angle $\theta$. The vector $\textbf{X}$ points from the common $\Gamma$ point of the mono-layer Brillouin zones to the center of the mini-Brillouin zone at the $K_+$ points.}\label{fig:BZ}
\end{figure}
\end{center}

For the Moir\'e Hamiltonian, we consider a commensurate Moir\'e pattern, obtained from an AA stacked bilayer where the top and bottom graphene layers are rotated relative to each other along an out-of-plane rotation axis centered at a hexagon by an angle $\theta$. This gives a Moir\'e super lattice with microscopic $C_{6z}$ symmetry, which is found to be a very good approximate low-energy symmetry even for microscopically less symmetric Moir\'e patterns obtained from different initial stacking alignments or different rotation axis \cite{Zou}. We choose to work with a commensurate pattern in order to use sharply defined Moir\'e lattice and reciprocal lattice vectors. However, the relevant properties of the electronic band structure around charge neutrality do not rely on the assumption of commensurability. In figure \ref{fig:BZ}(a) we show the mono-layer graphene Brillouin zone with our convention for the reciprocal lattice basis vectors and the high symmetry points $K_+$ and $K_-$.

We now consider following spinless (spin-orbit coupling is negligible) single-valley Moir\'e Hamiltonian

\begin{equation}
H(\textbf{k})=\sum_{\textbf{g}_1,\textbf{g}_2}\left(h^{++}(R(\theta/2)(\textbf{k}+\textbf{X}+\textbf{g}_1))\delta_{\textbf{g}_1,\textbf{g}_2} + h^{--}(R(-\theta/2)(\textbf{k}+\textbf{X}+\textbf{g}_1))\delta_{\textbf{g}_1,\textbf{g}_2} + \sum_{\tilde{\textbf{g}}}\left[ T_{\tilde{\textbf{g}}}^{+-} \delta_{\textbf{g}_1,\textbf{g}_2+\tilde{\textbf{g}}} + T_{\tilde{\textbf{g}}}^{-+} \delta_{\textbf{g}_1+\tilde{\textbf{g}},\textbf{g}_2} \right]\right) 
\end{equation}
Here, $\textbf{g}_1$ and $\textbf{g}_2$ lie on the Moir\'e reciprocal lattice, $R(\theta)$ is a rotation matrix over angle $\theta$, $h^{++}(\textbf{k})= th(\textbf{k})+\Delta_t\sigma^z$ ($h^{--}(\textbf{k})=th(\textbf{k})+\Delta_b\sigma^z$) is the mono-layer graphene Hamiltonian of the top (bottom) layer with hopping strength $t=2.61$ eV and a sublattice splitting $\Delta_t\sigma^z$ ($\Delta_b\sigma^z$). The mono-layer graphene Hamiltonian is given by
\begin{equation}
h(\textbf{k}) = \left(\begin{matrix} 0 & e^{i\textbf{k}\cdot\textbf{R}_A} +e^{i\textbf{k}\cdot(\textbf{R}_A-\textbf{R}_1)} + e^{i\textbf{k}\cdot(\textbf{R}_A-\textbf{R}_2)} \\   e^{-i\textbf{k}\cdot\textbf{R}_A} +e^{-i\textbf{k}\cdot(\textbf{R}_A-\textbf{R}_1)} + e^{-i\textbf{k}\cdot(\textbf{R}_A-\textbf{R}_2)} & 0 \end{matrix}\right)\, ,
\end{equation}
where $\textbf{R}_1,\textbf{R}_2$ are the graphene Bravais lattice vectors and $\textbf{R}_A,\textbf{R}_B$ are the sublattice vectors. $\textbf{X}$ is the position of the center of the mBZ at the mono-layer $K_+$-points as shown in Fig.\ref{fig:BZ}(b). In the commensurate case we are considering here, $\textbf{X}$ lies on the Moir\'e reciprocal lattice. The inter-layer coupling is given by the matrices

\begin{eqnarray}
T_{\textbf{0}} &=&\left(\begin{matrix} w_0 & w_1 \\ w_1& w_0 \end{matrix}\right)\\
T_{\textbf{g}_1}&=&\left(\begin{matrix} w_0 & w_1\omega \\ w_1\omega^*& w_0 \end{matrix}\right) \\
T_{\textbf{g}_2}&=&\left(\begin{matrix} w_0 & w_1\omega^* \\ w_1\omega & w_0 \end{matrix}\right)\, ,
\end{eqnarray}
where $\omega=e^{i2\pi/3}$, $\textbf{g}_1 = (R(\theta/2)-R(-\theta/2))\textbf{G}_1$ and $\textbf{g}_2 = (R(\theta/2)-R(-\theta/2))\textbf{G}_2$. The AB inter-layer hopping strength is $w_1= 98$ meV. To phenomenologically incorporate the corrugation of the bilayer system we have used an AA-AB inter-layer hopping ratio $w_0/w_1=0.85$.

\begin{center}
\begin{figure}[h]
a) 
\includegraphics[scale=0.35]{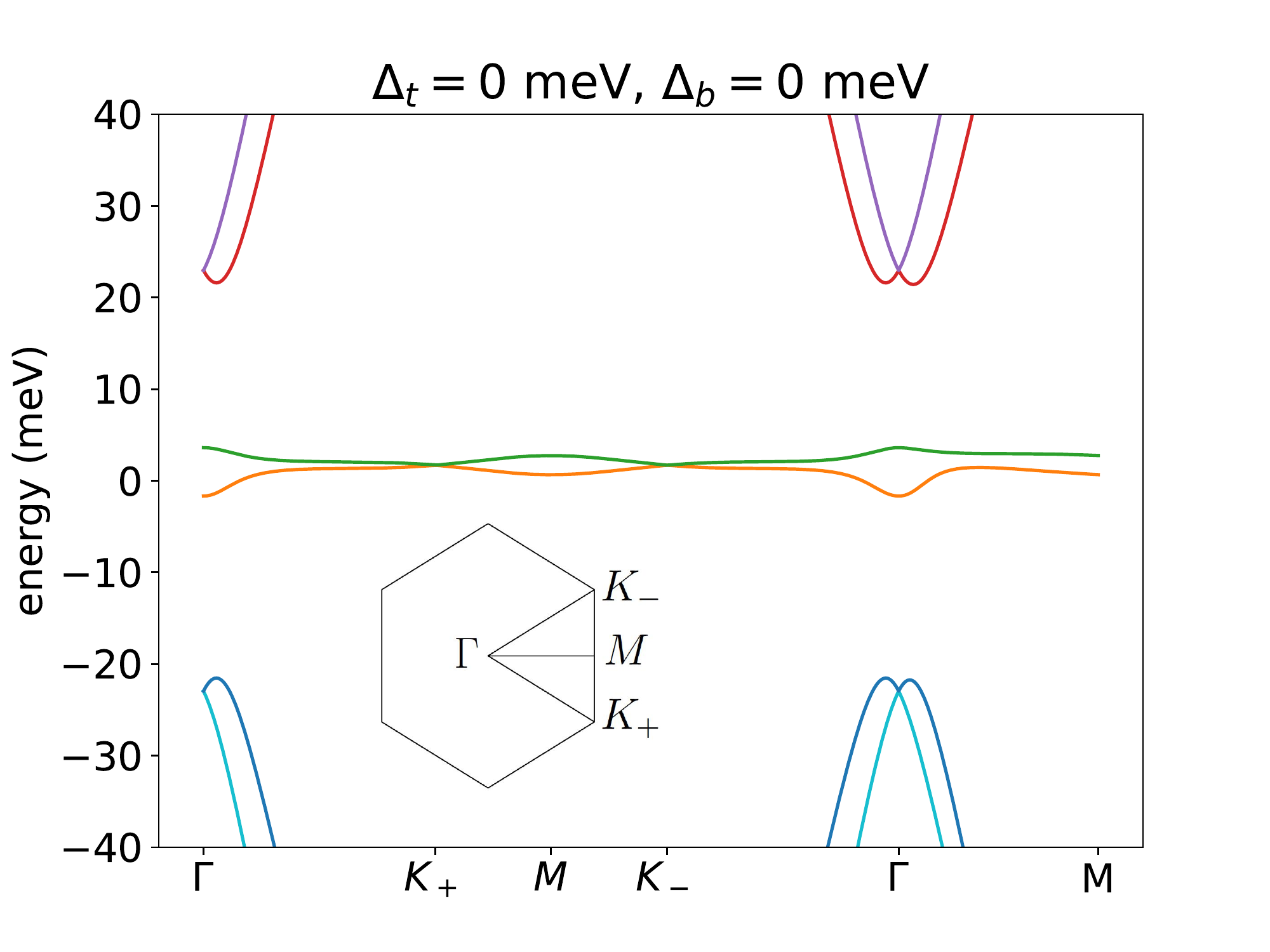} b)
\includegraphics[scale=0.35]{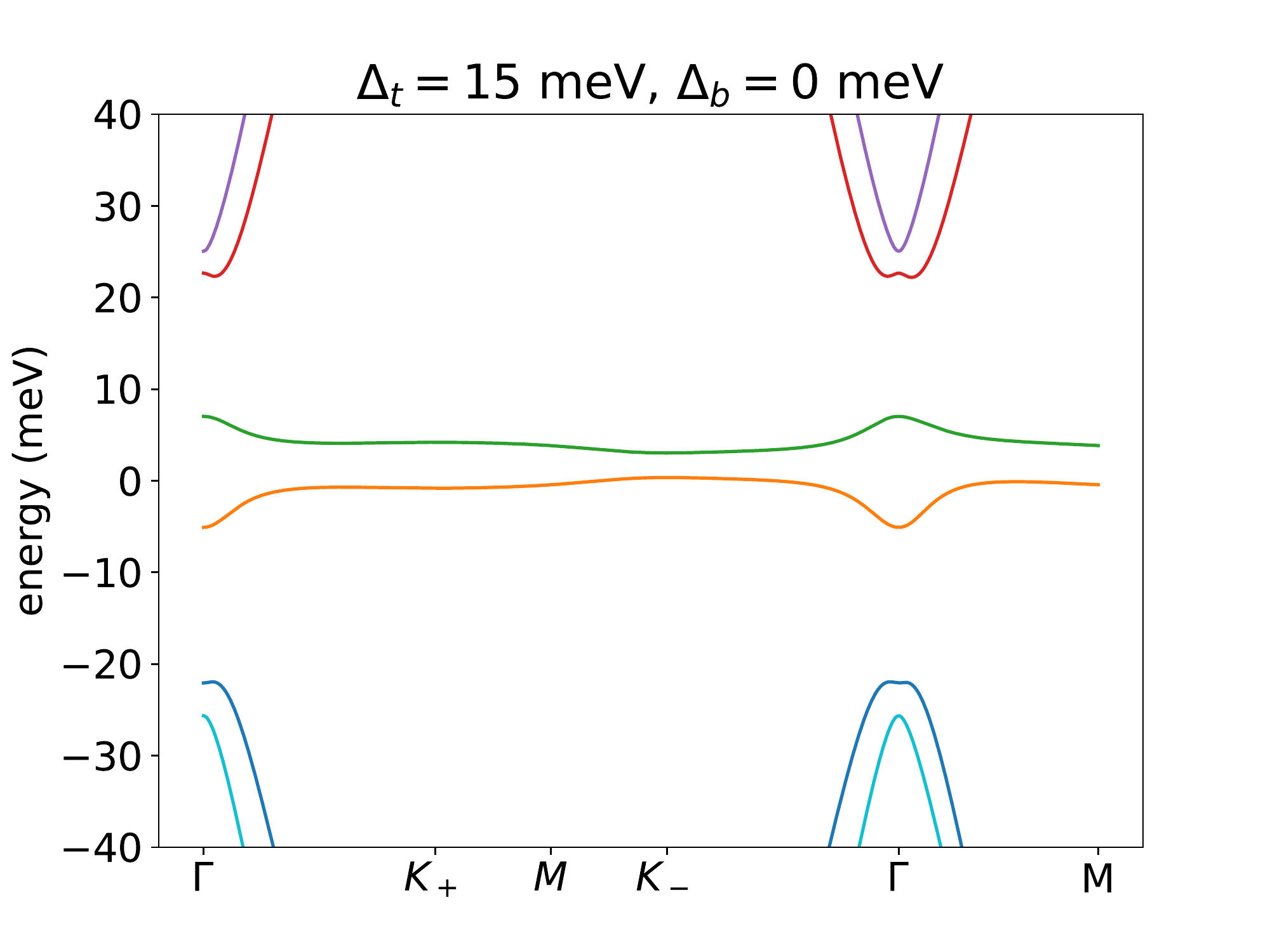} \\ c)
\includegraphics[scale=0.35]{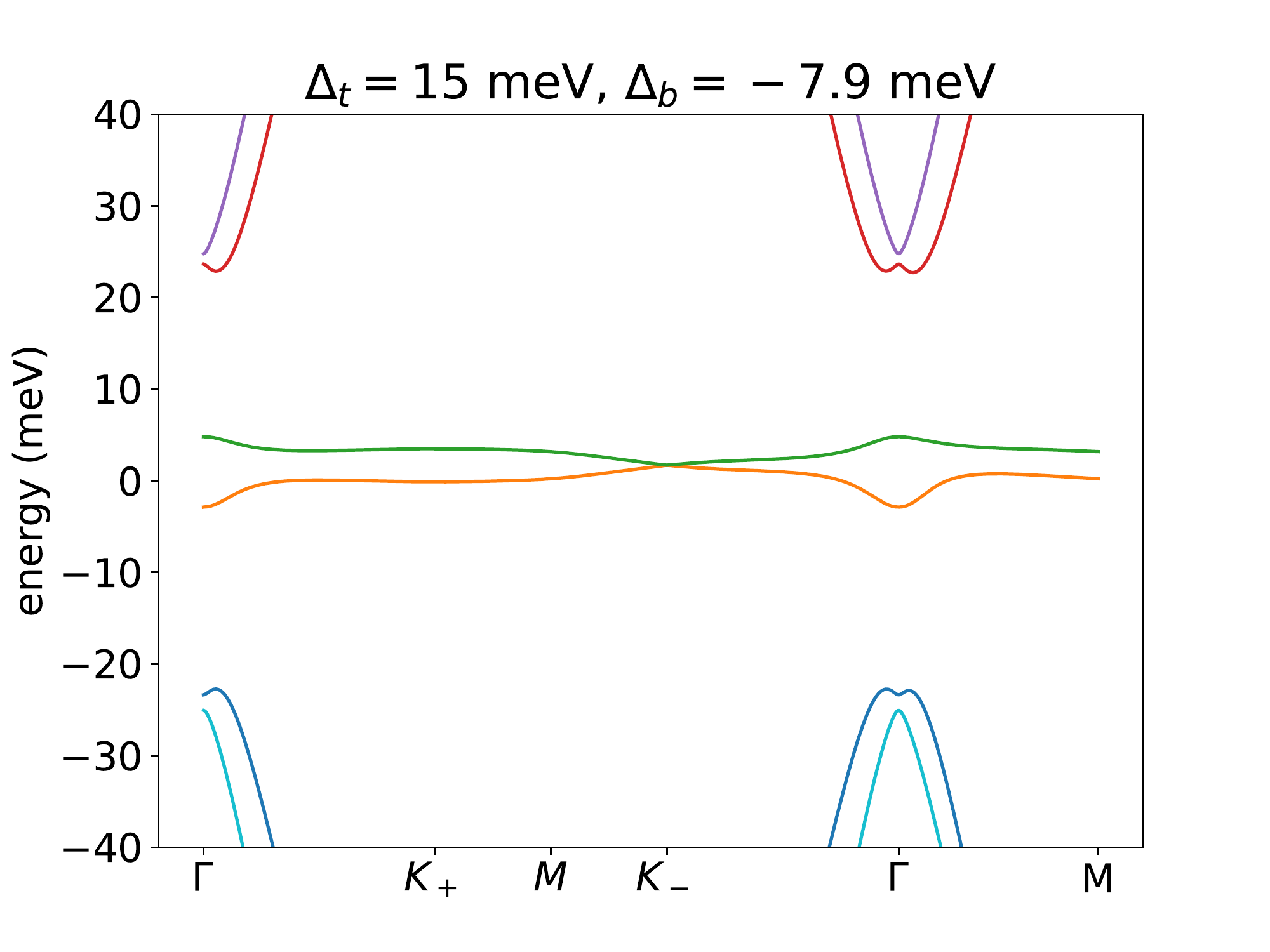} d)
\includegraphics[scale=0.35]{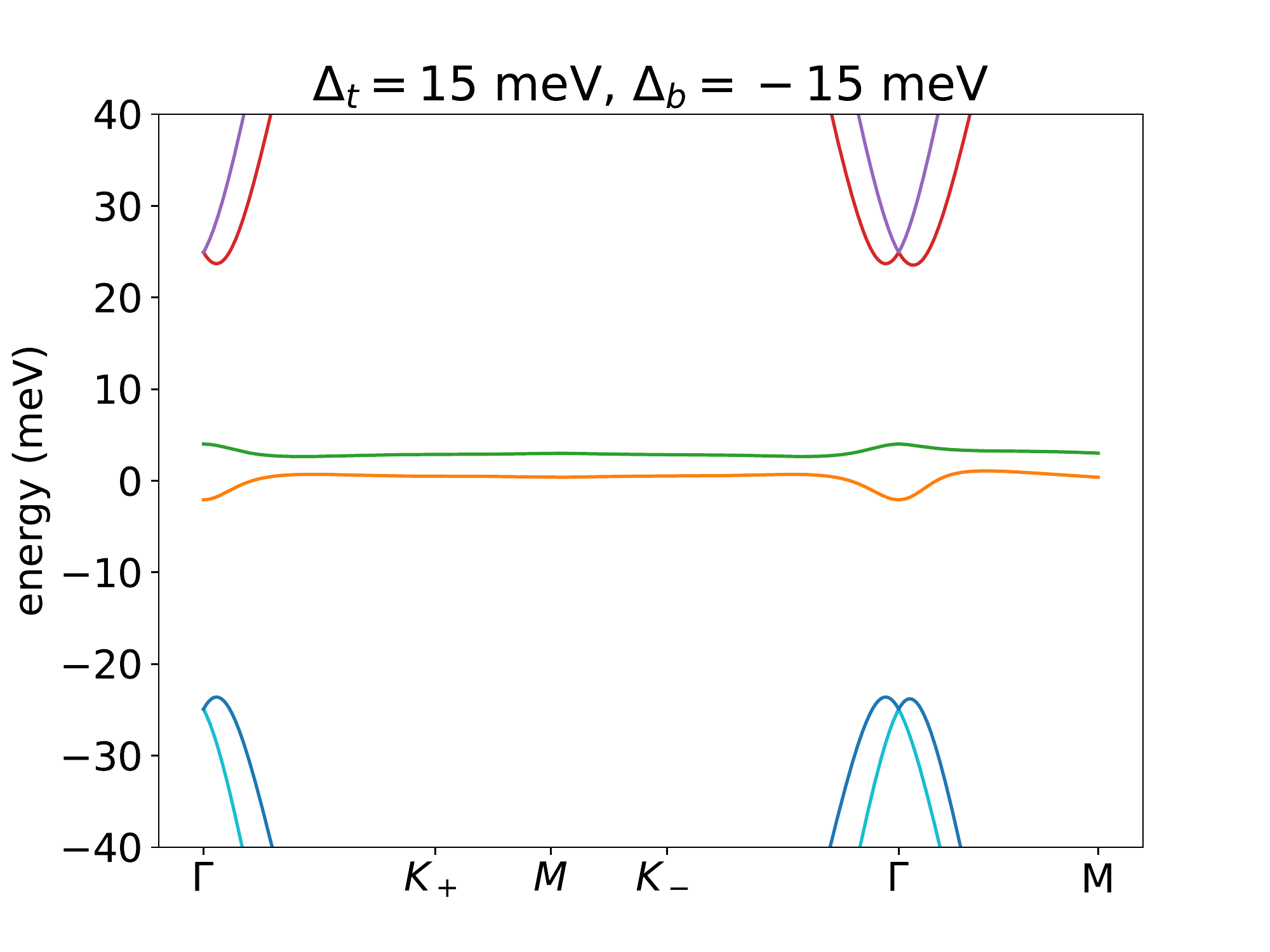} \\ e)
\includegraphics[scale=0.35]{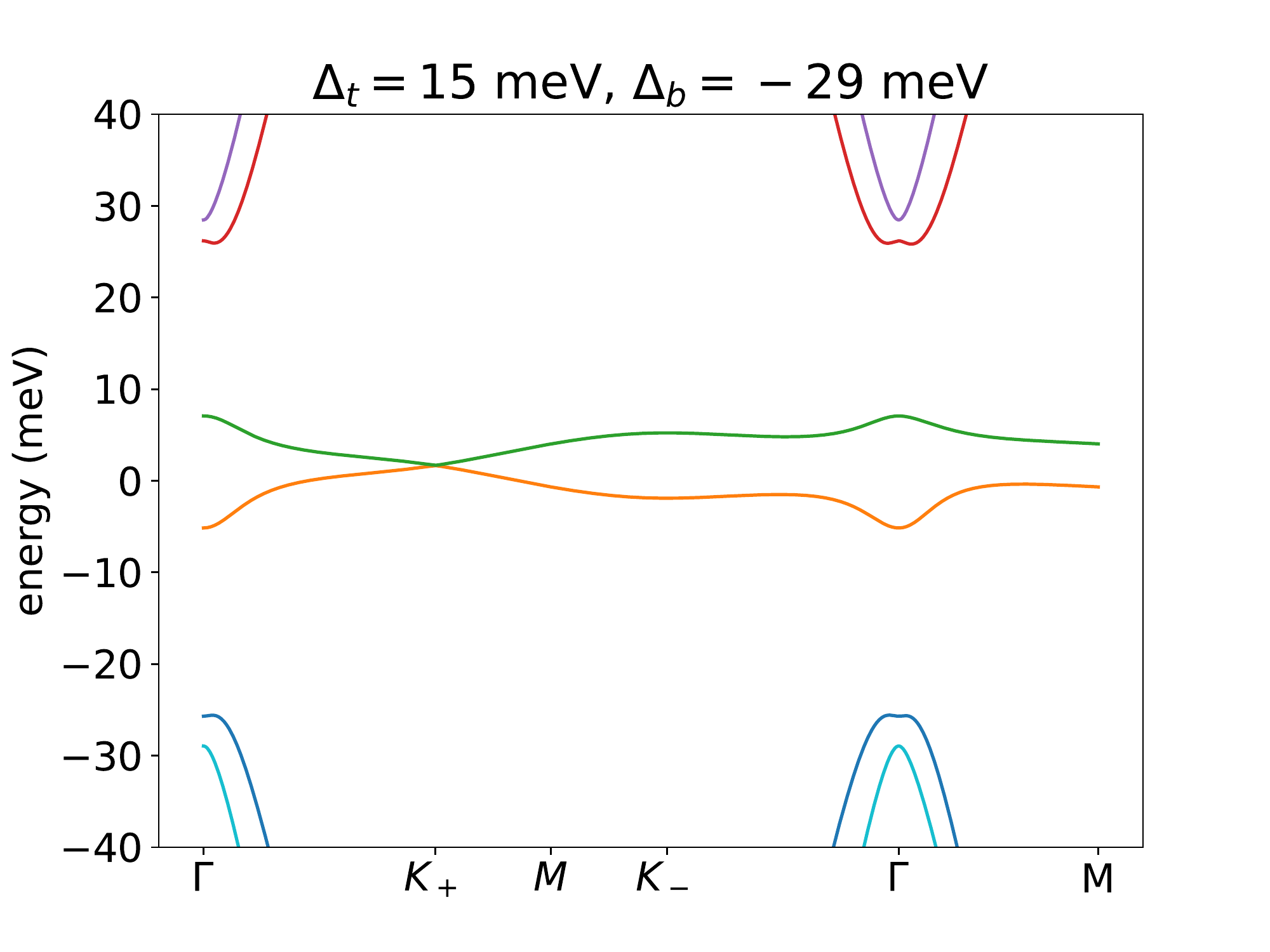} f)
\includegraphics[scale=0.35]{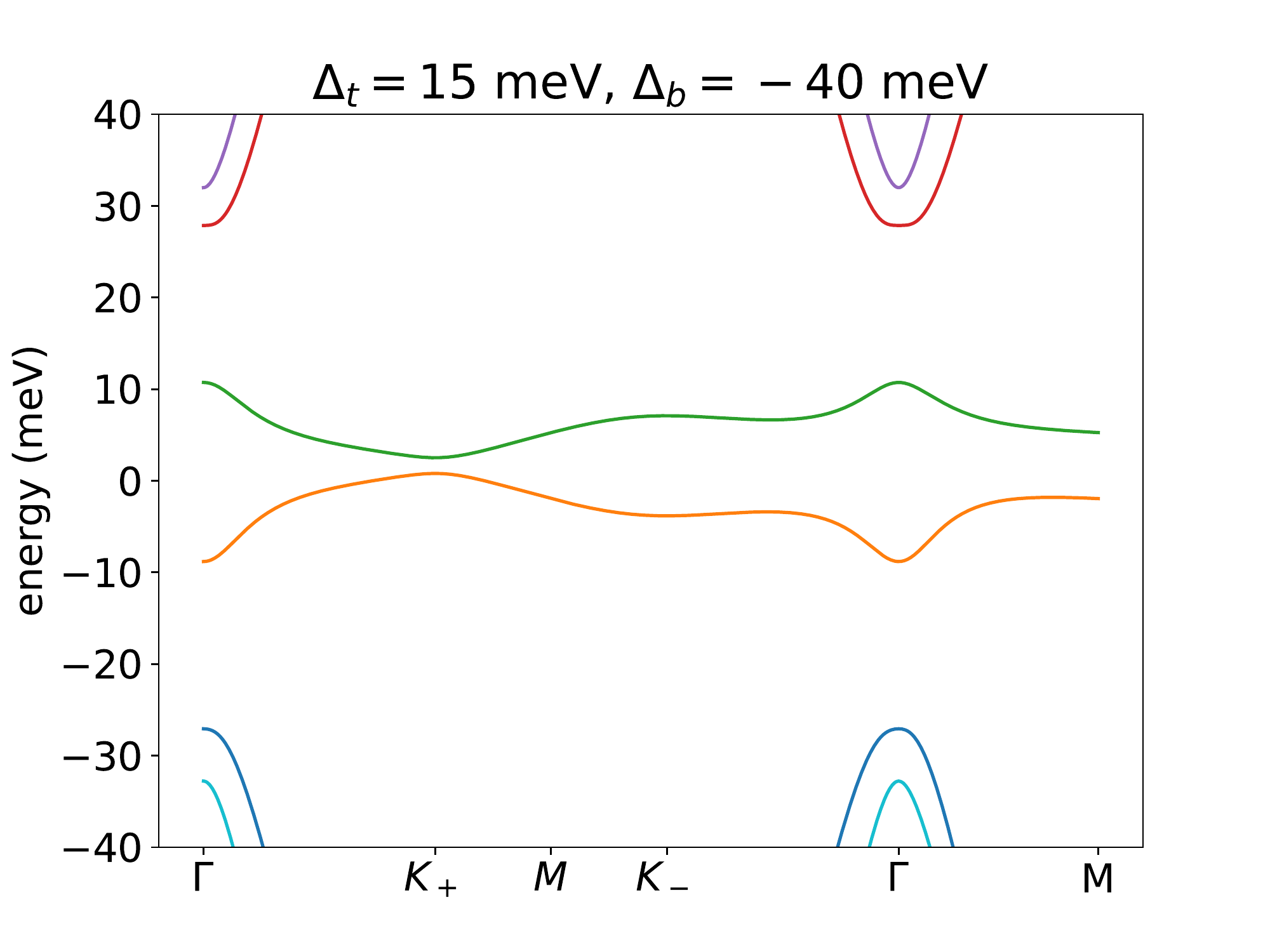}
\caption{Flat bands for the single-valley Moir\'e Hamiltonian at a twist angle $\theta\approx 1.05^\circ$ for different values of the sublattice splittings $\Delta_t$ and $\Delta_b$. The intralayer hopping is $t=2.61$ eV, the AA/BB inter-layer hopping $w_0=82$ meV and the AB/BA inter-layer hopping is $w_1=98$ meV. (a) Flat bands without sublattice splitting. The high-symmetry paths in the mBZ along which the band spectrum is shown are indicated. There are Dirac cones with small Fermi velocity at the $K_+$ and $K_-$ points of the mBZ. (b)-(f) Evolution of the flat bands for $\Delta_t=15$ meV and different values of $\Delta_b$. When $\Delta_b$ is zero, a non-zero $\Delta_t$ ensures that both Dirac cones acquire a mass. For decreasing negative values of $\Delta_b$, there are two Chern number changing transitions where a Dirac point occurs at either $K_+$ or $K_-$.}\label{fig:band}
\end{figure}
\end{center}

In Fig.~\ref{fig:band} we show the resulting flat bands in the mBZ around charge neutrality of the single-valley Moir\'e Hamiltonian along high-symmetry paths, for different strengths sublattice splittings $\Delta_t$ and $\Delta_b$. The twist angle in these calculations was $\theta \approx 1.05^\circ$. When $\Delta_t=\Delta_b=0$, the flat bands have Dirac cones at $K_+$ and $K_-$ and are separated from the dispersive bands by an energy gap of approximately $20$ meV. If one of the sublattice splittings is non-zero, both Dirac cones acquire a mass because of the strong inter-layer coupling. In Fig. \ref{fig:band} we also show different plots with $\Delta_t=15$ meV constant and decreasing negative $\Delta_b$ to show the two Chern number changing transitions where a Dirac cone occurs at either $K_+$ or $K_-$.

\section{Suppression of inter-valley scattering}

We write the single valley Moir\'e Hamiltonian schematically as
\begin{eqnarray}
H^\tau(\textbf{k}) & = & \sum_{\textbf{g}_1,\textbf{g}_2}\sum_{\xi,\sigma,\xi',\sigma'} |\textbf{k}+\textbf{g}_1,\xi,\sigma\rangle H^\tau_{(\xi,\sigma,g_1)(\xi',\sigma',g_2)}(\textbf{k})\langle \textbf{k}+\textbf{g}_2,\xi',\sigma'| \\
 & = & \sum_\mu \sum_{\textbf{g}_1,\textbf{g}_2}\sum_{\xi,\sigma,\xi',\sigma'} |\textbf{k}+\textbf{g}_1,\xi,\sigma\rangle U^\mu_{\tau,\textbf{k}}(\xi,\sigma,\textbf{g}_1)\epsilon^\tau_\mu(\textbf{k})U^{\mu*}_{\tau,\textbf{k}}(\xi',\sigma',\textbf{g}_2) \langle \textbf{k}+\textbf{g}_2,\xi',\sigma'|\, ,\nonumber
\end{eqnarray}
where again the vectors $\textbf{g}_i$ lie on the reciprocal lattice of the Moir\'e super lattice. Here, we introduced the notation that $\tau\in\{+,-\}$ represents the different Dirac valleys, located at the high symmetry $K-$points of the mono-layer graphene BZ. The sublattice degree of freedom is denoted by $\sigma\in \{A,B\}$, and the two graphene layers are labeled by $\xi\in\{+,-\}$. The carbon atoms are located at positions $\textbf{r}$, such that $\textbf{r}$ is of the form $\textbf{r}=R(\xi\theta/2)(m\textbf{R}_1+n\textbf{R}_2+\textbf{R}_{\sigma})$, where $m,n\in\mathbb{Z}$, $\textbf{R}_1,\textbf{R}_2$ are the graphene Bravais lattice vectors and $\textbf{R}_A,\textbf{R}_B$ the sublattice vectors. 

Importantly, for $H^\tau(\textbf{k})$ we define the momentum $\textbf{k}$ relative to the center of the mini-Brillouin zone located at the mono-layer $K_\tau$-points of top and bottom layer. In the second line we diagonalized the Moir\'e Hamiltonian using the unitary matrices $U$. Because we are interested in one band per valley, we drop the $\mu$ band index and associate $\tau$ with the band label. In this notation, we write the flat band states in each valley as
\begin{eqnarray}
c_{\textbf{k},\tau,s}^\dagger & = & \sum_{\textbf{g}}\sum_{\xi,\sigma}U_{\tau,\textbf{k}}(\xi,\sigma,\textbf{g}) \psi^\dagger_{\textbf{k}+\textbf{g}+\tau\textbf{K},\xi,\sigma,s} \label{notation1}\\
 & = &  \sum_{\textbf{r}}\sum_{\textbf{g}} U_{\tau,\textbf{k}}(\xi,\sigma,\textbf{g})e^{i(\textbf{k}+\textbf{g}+\tau\textbf{X})\cdot\textbf{r}}\psi^\dagger_{\textbf{r},s}\label{notation2}\\
 & \equiv & \sum_{\textbf{r}} \phi_{\tau,\textbf{k}}([\textbf{r}])e^{i(\textbf{k}+\tau\textbf{X})\cdot\textbf{r}} \psi^\dagger_{\textbf{r},s} \\
 & \equiv & \sum_{\textbf{r}}\tilde{\phi}_{\tau,\textbf{k}}([\textbf{r}])e^{i\textbf{k}\cdot\textbf{r}} \psi^\dagger_{\textbf{r},s}\,.
\end{eqnarray}
Here we have introduced the notation $\textbf{r}=[\textbf{r}]+\textbf{t}$, where $[\textbf{r}]$ is the part of $\textbf{r}$ lying in the Moire unit cell centered at the origin and $\textbf{t}$ are Moire lattice vectors. To clarify the notation in going from Eq. \eqref{notation1} to Eq. \eqref{notation2}, recall that the position label $\textbf{r}$ contains the information contained in the labels $\xi$ and $\sigma$ via the relation $\textbf{r}=R(\xi\theta/2)(m\textbf{R}_1+n\textbf{R}_2+\textbf{R}_{\sigma})$. The index $s$ refers to spin. Inverting the above expression now gives us the electron raising operator at position $\textbf{r}$ projected in the two flat bands:

\begin{eqnarray}
\psi^\dagger_{\textbf{r},s} & = &  \sum_{\textbf{k},\tau} e^{-i\textbf{k}\cdot\textbf{r}} \tilde{\phi}^*_{\tau,\textbf{k}}([\textbf{r}]) c^\dagger_{\textbf{k},\tau,s} 
\end{eqnarray}
With the projected electron raising operator, which we simply rewrite as $\psi^\dagger_{\textbf{t},[\textbf{r}],s}$, we now define

\begin{eqnarray}
n_{[\textbf{r}]}(\textbf{q}) & = & \sum_{s}\sum_{\textbf{t}} e^{-i\textbf{q}\cdot\textbf{t}}\psi^\dagger_{\textbf{t},[\textbf{r}],s}\psi_{\textbf{t},[\textbf{r}],s} \\
& = & \sum_s \sum_{\tau,\tau'} \sum_{\textbf{k}} \phi^*_{\tau,\textbf{k}}([\textbf{r}]) \phi_{\tau',\textbf{k}+\textbf{q}}([\textbf{r}]) e^{i(\textbf{q}+(\tau'-\tau)\textbf{X})\cdot[\textbf{r}]} c^\dagger_{\textbf{k},\tau,s}c_{\textbf{k}+\textbf{q},\tau',s}
\end{eqnarray}
Using the above expression we can write the microscopic Coulomb interaction projected into the flat bands as

\begin{eqnarray}
H^{i} & = &   \sum_{\textbf{q}}\sum_{[\textbf{r}],[\textbf{r}']}V_{\textbf{q}}([\textbf{r}]-[\textbf{r}']):n_{[\textbf{r}]}(\textbf{q})n_{[\textbf{r}]}(-\textbf{q}): \\
 & = & \sum_{\textbf{q}}\sum_{s,s'}\sum_{\tau_1,\tau'_1,\tau_2,\tau'_2} \sum_{\textbf{k},\textbf{k}'} V^{\textbf{k},\textbf{k}'}_{\tau_1,\tau'_1,\tau_2,\tau'_2}(\textbf{q}) c^\dagger_{\textbf{k},\tau_1,s} c^\dagger_{\textbf{k}',\tau_2,s'}c_{\textbf{k}'-\textbf{q},\tau'_2,s'} c_{\textbf{k}+\textbf{q},\tau'_1,s}\, ,
\end{eqnarray}
where
\begin{equation}
V_{\textbf{q}}([\textbf{r}]-[\textbf{r}']) = \sum_{\textbf{t}} e^{i\textbf{q}\cdot\textbf{t}} V(\textbf{t}+[\textbf{r}]-[\textbf{r}'])\, ,
\end{equation}
and $V(\textbf{r}-\textbf{r}')$ is the microscopic Coulomb interaction. The projected interaction coefficients are given by
\begin{equation}
V^{\textbf{k},\textbf{k}'}_{\tau_1,\tau'_1,\tau_2,\tau'_2}(\textbf{q}) = \sum_{[\textbf{r}],[\textbf{r}']} \phi^*_{\tau_1,\textbf{k}}([\textbf{r}]) \phi_{\tau'_1,\textbf{k}+\textbf{q}}([\textbf{r}])e^{i(\tau_2'-\tau_2)\textbf{X}\cdot[\textbf{r}]} e^{-i\textbf{q}\cdot([\textbf{r}]-[\textbf{r}'])}V_{\textbf{q}}([\textbf{r}]-[\textbf{r}'])\phi^*_{\tau_2,\textbf{k}'}([\textbf{r}']) \phi_{\tau'_2,\textbf{k}'-\textbf{q}}([\textbf{r}'])e^{i(\tau_1'-\tau_1)\textbf{X}\cdot[\textbf{r}]'} 
\end{equation}
Now it is important to remember that $\phi_{\tau,\textbf{k}}(\textbf{r}) = \sum_{\textbf{g}} U_{\tau,\textbf{k}}(\xi,\sigma,\textbf{g})e^{i\textbf{g}\cdot\textbf{r}}$. Because for the flat band states $U_{\tau,\textbf{k}}(\xi,\sigma,\textbf{g})$ decays fast with $|\textbf{g}|$, $\phi_{\tau,\textbf{k}}(\textbf{r})$ varies slowly within the Moir\'e unit cell. So if $V(\textbf{r}-\textbf{r}')$ is sufficiently long-range (like Coulomb), then the sums over $[\textbf{r}]$ and $[\textbf{r}']$ will suppress the terms with $\tau_1\neq \tau'_1$ and $\tau_2\neq \tau'_2$. For this reason, we restrict to the dominant density-density terms in our effective Landau-level problem.

\section{Construction of Wannier-Qi states}

Because the single-valley flat bands split by the one-sided staggered potential have Chern number $\pm 1$, one cannot construct exponentially localized Wannier functions for these bands \cite{Thouless}. However, using the right gauge choice it is possible to construct Wannier functions that are exponentially localized along one direction. Using these quasi-one dimensional Wannier states there exists a natural mapping from the lattice system to a Landau-level system, as pointed out by Qi in Ref.\cite{Qi}. Here we review this mapping in the context of TBG.

Consider a system with periodic boundary conditions along two directions, which we refer to as the $x$ and $y$-directions. Using the flat band states as defined in the previous appendix, we construct the superlattice Wannier-Qi functions as follows:
\begin{eqnarray}
d^\dagger_{x_0,k_y,\tau,s} & = & \sum_{k_x} e^{-ix_0k_x}  e^{i\alpha_\tau(\textbf{k})}c^\dagger_{\textbf{k},\tau,s} \\ 
 & = & \sum_{\textbf{r}}\sum_{k_x} e^{i\alpha_\tau(\textbf{k})}\tilde{\phi}_{\tau,\textbf{k}}([\textbf{r}]) e^{-ix_0k_x}e^{i\textbf{k}\cdot\textbf{r}}\psi^\dagger_{\textbf{r},s} \\
 & = & \sum_{[\textbf{r}]}\sum_{\textbf{t}'}\left( \sum_{k_x}e^{-i(x_0-t_x')k_x} e^{i\alpha_\tau(\textbf{k})} \tilde{\phi}_{\tau,\textbf{k}}([\textbf{r}])\right) e^{ik_y t'_y} \psi^\dagger_{[\textbf{r}]+\textbf{t}',s} \\
 & \equiv & \sum_{[\textbf{r}]}\sum_{\textbf{t}'} W_{\tau,x_0,k_y}(\textbf{r}) e^{ik_yt'_y}\psi^\dagger_{\textbf{r},s}\, .
\end{eqnarray}
Here, $e^{i\alpha_\tau(\textbf{k})}$  ensures an optimal gauge choice such that the functions $W_{\tau,x_0,k_y}(\textbf{r})$ are exponentially localized in the $x$-direction around the lattice position $x_0$. We now imagine adiabatically threading $2\pi$ flux through the hole of the torus, such that the flux is felt by a particle moving on closed path in the $y$-direction. Because of the Chern number $|C|=1$, this adiabatic process will change the polarization in the $x$-direction by one `polarization quantum' \cite{Resta,Vanderbilt}, which means that the centers of the Wannier functions all shift by one Moire lattice vector along the $x$-axis (the direction in which they shift depends on the sign of the Chern number). This implies that $W_{\tau,x_0,k_y+g}(\textbf{r}) = W_{\tau,x_0+\tau t_x,k_y}(\textbf{r})$, where $g$ is the norm of the Moir\'e reciprocal basis vectors. Therefore, we can use $k\equiv k_y + \tau g $ as a single label for our Wannier-Qi states $W_{\tau,k}(\textbf{k})$ (for each $k_y \in [0,g]$, there is one Wannier function with a particular value for $x_0$ in each Moir\'e unit cell). One can now straightforwardly map the Chern band to a LLL, by replacing each Wannier-Qi function $W_{\tau,k}(\textbf{r})$ by the corresponding LLL Gaussian wave function. One of the main approximations in using the LLL states instead of the Wannier-Qi states of the twisted bilayer is that we ignore any Berry-curvature inhomogeneity. 

\section{Exciton vortex lattice in the lowest Landau level}
The perpendicular magnetic field seen by the exciton order parameter $\Delta(\r) = \langle c^\dagger_{+,\r} c_{-,\r} \rangle$ induces a vortex lattice in the order parameter. Since $\Delta(\r)$ essentially behaves like a charge $q = 2e$ object in a magnetic field, the solution to this vortex lattice may be obtained by solving the Ginzburg Landau (GL) equation for $\Delta(\r)$. For analytical tractability, we focus on vortex lattice solution of the linearized GL equation. Our solution is exact only at the upper critical field $H_{c2}$ of the corresponding superconductor, but we expect our results to be valid more generally. In this limit, the problem reduces to the solving the Schrodinger equation for a single particle of charge $2e$. In the Landau gauge $\A = B x \hat{y}$, this solution is given by
\beq
\Delta(\r) = \sum_k C_k e^{i k y} e^{- \frac{1}{2\xi^2}(x - k \xi^2)^2}, ~~~ \xi = \frac{l_B}{\sqrt{2}}
\eeq

The exciton vortex lattice we consider has $2\pi$ flux through each plaquette of the square lattice of side $a$, i.e, $a^2 = 2 \pi l_B^2$. Since each elementary vortex carries a flux of $\pi$, we therefore expect two elementary vortices within a plaquette. Inspired by the computation of similar vortex patterns for the superconducting order parameter in Ref.~\cite{Mishmash}, we choose the vortex lattice wavefunction to be symmetric under magnetic translations $\T_1 = \T(a \hat{y})$ and $\T_2 = \T\left(\frac{a}{2}(\hat{x} + \hat{y})\right)$. Note that the magnetic translation operators for a particle of charge $q$ in a magnetic field $\mathbf{B} $ satisfy the following algebra:
\beq
\T_\R \, \T_{\R^\prime} = e^{i q \mathbf{B} \cdot (\R \times \R^\prime)/ \hbar} \, \T_{\R^\prime} \, \T_\R
\label{eq:MagAlg}
\eeq
Since $\Delta(\r)$ is a charge $q = 2e$ order parameter, we have $q \mathbf{B} \cdot (\R_1 \times \R_2)/\hbar = (2 e) B (a^2/2)/\hbar = 2 \pi $ implying that $\T_1$ and $\T_2$ commute with each other. Being magnetic translation operators, they commute with the free Hamiltonian of a particle of charge $q$. Since our goal is to express $\Delta(\r)$ in the Bloch basis, where the eigenstates of particles with charge $q = e$ are invariant under the square lattice translations $\T_1 = \T(a \hat{y})$ and $\T_3 = \T(a \hat{x})$, we also choose the phases of $\T_1$ and $\T_2$ such that $\T_3 = \T_2^2 \T_1^{-1}$ is identically satisfied. Consistent with these conditions, we find that
\beq
\T_1 = e^{- i a p_y/\hbar } \; ; ~~~ 
\T_2 = e^{i \pi/2} e^{ i a y/l_B^2 } e^{- i a(p_x + p_y)/(2 \hbar)}  
\eeq
Now we impose the magnetic translation symmetry requirements on $\Delta(\r)$. For $\T_1$, we have
\beq
\T(a \hat{y}) \Delta(\r) = \sum_k C_k e^{-i k a} e^{i k y} e^{- \frac{1}{2\xi^2}(x - k \xi^2)^2} = \Delta(\r) \implies k = k_j = \frac{2 \pi j}{a} = j Q \text{ for } j \in \mathbb{Z}
\eeq
where we have defined $Q = 2\pi/a$. Therefore, we can write 
\beq
\Delta(\r) = \sum_{j = -\infty}^{\infty} C_j e^{i k_j y} e^{- \frac{1}{2\xi^2}(x - k_j \xi^2)^2}
\eeq
For $\T_2$, we have, using $2 \pi l_B^2 = a^2$ or equivalently $a = Q l_B^2$,
\beq
\T_2 \Delta(\r) &=& \sum_{j = -\infty}^{\infty} C_j  e^{i \pi/2} e^{- i k_j a} e^{i (k_j + a/l_B^2) y} e^{- \frac{1}{2\xi^2}(x - (k_j+a/l_B^2) \xi^2)^2} =  \sum_{j = -\infty}^{\infty} C_j ~ e^{i \pi j + i \pi/2 } ~ e^{i k_{j+1}y} e^{- \frac{1}{2\xi^2}(x - k_{j+1} \xi^2)^2} \nn
&=&  \Delta(\r) \implies C_j ~ e^{i \pi j + i \pi/2 } = C_{j+1} \implies C_j = e^{i \frac{\pi}{2} j^2} C_0 
\eeq
Therefore, we have the following form of $\Delta(\r)$:
\beq
\Delta(\r) = C_0 \sum_{j = -\infty}^{\infty} e^{i \frac{\pi}{2} j^2} e^{i k_j y} e^{- \frac{1}{2\xi^2}(x - k_j \xi^2)^2}
\eeq
We can now find the projection of $\Delta(\r)$ on the single particle Bloch wave-functions. We focus on the lowest Landau level since we are only interested in $C = \pm 1$ bands. We define
\beq
\Delta_{00}(\k,\k^\prime) =  \int d\r \; \Delta(\r) \phi_{+,\k}^*(\r) \phi_{-,\k^\prime,}(\r)
\eeq
where $\phi_{\pm, \k}(\r)$ are the Bloch wave-functions defined in Eq.~(1) in the main text. Given the  symmetry of $\Delta(\r)$ under magnetic translations $\T_{1}$ and $\T_3$, we expect it to be diagonal in Bloch space. Indeed, we find that $\Delta_{00}(\k,\k^\prime) = \Delta_{\k} \delta_{\k,\k^\prime}$, where
\beq
\label{eq:VorMom}
\Delta_\k = \Delta_0 \, \sum_{j = -\infty}^{\infty} e^{-i \frac{\pi}{2} j^2}  e^{- \frac{1}{4}(2 k_y + j Q)^2 l_B^2}  e^{- i k_x (2 k_y + j Q )l_B^2}
\eeq
where $\Delta_0 = \frac{C_0}{\sqrt{2}}$ is a measure of the overall strength of the exciton vortex lattice order parameter. 

We can recast $\Delta_\k$ in terms of the Jacobi theta function as follows
\beq
\Delta_\k & = & \Delta_0 e^{ (k_x - i k_y)^2 \l_B^2 - k_x^2 l_B^2} \; \vartheta_3\left(z = - \frac{k_x - i k_y}{Q}; \, \tau = \frac{e^{-i \pi/4}}{\sqrt{2}} \right)  \nn
&& \text{ with } \vartheta_3(z;\tau) = \sum_{n = - \infty}^{\infty} e^{i \pi \tau n^2 + i 2 \pi n z }\, .
\eeq
The Jacobi theta function has zeros at $z = m + n \tau + 1/2 + \tau/2$, where $m,n \in \mathbb{Z}$. Restricting to the first BZ, we find that $\Delta_{\k} = 0$ at $\k = \pm \k_0$, with $\k_0 = (\pi/2,-\pi/2)$. Further, a power series expansion about the zeros shows that $\Delta_{\pm \k_0 + \q} = \pm A (q_x - i q_y) + O(q^2)$ (for some $A \in \mathbb{C}$), indicating that both nodes have the same chirality. The presence of these two nodes in the BZ, which is a topological requirement of $\Delta_\k$ arising from hybridization of bands with $C = \pm 1$, is confirmed by plotting the absolute value of $\Delta_\k$ in Fig. 2 in the main text.

The nodes in the exciton order parameter are intimately related to the Dirac cones of the $C_{2z}T$-symmetric single-valley Moir\'e Hamiltonian, and the associated Wannier obstruction \cite{Po,Zou,Po2,Bernevig,Yuan,Kang,Koshino}. To see this, consider a free fermion Hamiltonian with two bands that are isolated from the other bands, such that the momentum space Hamiltonian projected onto the two isolated bands takes the form
\begin{eqnarray}
H(\textbf{k})\big|_{+,-} &= &  (\epsilon_{\textbf{k}}-\mu)|u_{+,\textbf{k}}\rangle\langle u_{+,\textbf{k}}| - (\epsilon_{\textbf{k}}-\mu)|u_{-,\textbf{k}}\rangle\langle u_{-,\textbf{k}}| \nonumber\\ 
& & + \Delta_{\textbf{k}} |u_{+,\textbf{k}}\rangle\langle u_{-,\textbf{k}}|+\Delta^*_{\textbf{k}}|u_{-,\textbf{k}}\rangle\langle u_{+,\textbf{k}}|\, ,
\end{eqnarray}
where $|u_{\pm,\textbf{k}}\rangle$ are the periodic parts of the Bloch states in the two-band subspace. The dispersion of the two bands is given by $\pm \sqrt{(\epsilon_{\textbf{k}}-\mu)^2+|\Delta_{\textbf{k}}|^2}$ (for simplicity, but without loss of generality, we use a particle-hole symmetric projected Hamiltonian). We consider the situation where $|u_{\pm,\textbf{k}}\rangle$ has Chern number $\pm 1$, i.e. $\frac{1}{2\pi}\int_{\textbf{k}} \nabla\times \textbf{A}_\pm = \pm 1$,
where $\textbf{A}_\pm = -i\langle u_{\pm,\textbf{k}}|\nabla|u_{\pm,\textbf{k}}\rangle$. Using the same reasoning as for the exciton order parameter, and the fact that the Hamiltonian has to be periodic over the Brillouin zone, we conclude that $\Delta_{\textbf{k}}$ has two zeros in the Brillouin zone around which its phase winds by $2\pi$. Now imagine tuning $\mu$ from minus infinity to plus infinity. In this process the Chern number of the lowest energy band changes from $+1$ to $-1$, which is only possible if the energy gap between the two bands closes for intermediate values of $\mu$. From the band dispersion, we see that this gap closing will occur precisely at the momenta where the zeros of $\Delta_{\textbf{k}}$ are located. At these points, the nodes in $\Delta_{\textbf{k}}$ give rise to Dirac cones with the same chirality.

\section{Analytical Hartree-Fock energetics for the lowest Landau level model}
In this section, we compute the Hartree-Fock (HF) energy in a variational Slater determinant state for the Hamiltonian $H = H^i + H^p$, which we recall for completeness. 
\beq
H^p &=& \sum_{\k, \tau} \epsilon_\k \, c^\dagger_{\tau,\k}c_{\tau,\k} \, , \text{  where } \epsilon_\k = - \frac{W}{4}\left[ \cos\left( \frac{2\pi x}{a}\right) + \cos\left( \frac{2\pi y}{a}\right) \right] \nn
H^i  &=&  \frac{1}{2 N_\phi},   \sum_{\q,\tau,\tau'}V_{\tau,\tau'}(\q):n_\tau(\q) n_{\tau'}(-\q): \, , \text{ where } V_{\tau,\tau'}(\q) = u_0(\textbf{q})\left(\begin{matrix} 1 & 1 \\ 1 & 1  \end{matrix}\right) + u_1(\textbf{q}) \left(\begin{matrix} 1 & -1 \\ -1 & 1  \end{matrix}\right) 
\label{eq:AppH}
\eeq
In Eq.~(\ref{eq:AppH}), $u_0(\q)$ is the symmetric part of the interaction, while $u_1(\q)$ represents the anisotropy. We evaluate $\langle H \rangle$ for a variational Slater determinant state, which can capture the fully valley polarizd Chern insulator, the unpolarized and partially polarized metallic states, and the exciton condensate (both uniform and a vortex lattice) in different limits. Our variational state $\ket{\psi_{MF}}$ may be taken to be the Slater determinant ground state of a mean field Hamiltonian of the form $H_{MF} = \sum_{\k,\tau,\tau^\prime} c^\dagger_{\k,\tau} h_{\tau,\tau^\prime}(\k) c_{\k,\tau^\prime}$, where 
\beq
h_{\tau,\tau^\prime}(\k) = \begin{pmatrix}
\varepsilon_{\k} + h & \Delta^*_{\k} \\
\Delta_{\k} & \varepsilon_{\k} - h
\end{pmatrix} 
\eeq
Such a state $\ket{\psi_{MF}}$ is characterized by two variational parameters, the polarization $P_v$ (determined by $h$) and the strength of the excitonic order parameter $\Delta_0$. In the $|h| \gg \text{max}\{|\varepsilon_{\k}|\}$ and $\Delta_{\k} = 0$ limit, our Slater determinant state is thus fully valley polarized with $|P_v| = 1$, while for $h = \Delta_{\k} = 0$ we have an unpolarized metal. For $h \neq 0$ and $\Delta_{\k} = 0$, the state is a partially polarized metal with different chemical potential for the $\tau = \pm$ valleys. For $h = 0$ and $\Delta_\k \neq 0$, the state is  an excitonic condensate or vortex lattice (depending on the precise structure of $\Delta_\k$), with $P_v = 0$ fixed by the discrete $z \rightarrow -z$ symmetry of $H_{MF}$. The most general state will have both $h$ and $\Delta_\k$ non-zero. 

Evaluating the covariance matrix for $\ket{\psi_{MF}}$ with a chemical potential $\mu$ which fixes the filling of the mean-field bands gives:
\beq
\label{eq:Pmat}
\langle c^\dagger_{\tau, \k} c_{\tau^\prime \k^\prime} \rangle &=& P_{\tau, \tau^\prime}(\k) \, \delta_{\k,\k^\prime}, \text{ with } 
P_{\tau, \tau^\prime}(\k)  =  \begin{pmatrix} 
|u_\k|^2 \Theta_{\k,\alpha} + |v_\k|^2 \Theta_{\k,\beta} & u_\k v_\k (\Theta_{\k,\alpha} - \Theta_{\k,\beta}) \\
u_\k^* v_\k^* (\Theta_{\k,\alpha} - \Theta_{\k,\beta}) & \; \; |v_\k|^2 \Theta_{\k,\alpha}+ |u_\k|^2 \Theta_{\k,\beta}
\end{pmatrix} \nn
\text{ where } u_{\k} &=& \cos\left(\frac{\theta_\k}{2}\right),~ v_\k = e^{i \phi_\k} \sin\left(\frac{\theta_\k}{2}\right),~ \tan(\theta_\k) = \frac{|\Delta_\k|}{h},~ e^{i \phi_\k} = \frac{\Delta_\k}{|\Delta_\k|} \text{ and } \Theta_{\k,\alpha(\beta)} = \Theta(\mu - E_{\k,\alpha (\beta)})
\eeq
One can indeed check that $P_{\tau, \tau^\prime}(\k)$ is a projector matrix, i.e, $P^2 = P$, as expected for a Slater determinant state. We use these correlators to evaluate $\langle H \rangle$ via Wick's theorem.
 \beq
 \lim_{N_\phi \rightarrow \infty} \frac{E^{HF}}{N_\phi} &=& \frac{1}{2}  \int_{\k, \k^\prime,\q}  \sum_{\tau,\tau^\prime} V^{LL}_{\tau \tau^\prime}(\q) \bigg( P_{\tau,\tau}(\k) P_{\tau^\prime, \tau^\prime} (\k^\prime) \delta_{\q,0} - P_{\tau,\tau^\prime}(\k - \q/2) P_{\tau^\prime, \tau} (\k + \q/2) \delta_{\k,\k^\prime}\bigg) + \int_{\k} \varepsilon_\k \Tr(P(\k)) \nn
 \label{eq:EHF}
\eeq
where we have taken the thermodynamic limit, set the lattice spacing $a = 1$ and used the notation $\int_{\k} = \int \frac{d^2k}{(2\pi)^2}$ to denote integration over the first BZ. We now focus on different limits where we can analytically compute the regularized energy density $e^{HF}(P_v,\Delta_0) \equiv \lim_{N_\phi \rightarrow \infty} \left(  \frac{E^{HF}}{N_\phi} - \frac{u_0(\textbf{0})}{2} \right) $ (where we have subtracted the formally infinite self-energy contribution that is canceled by the positive background) and get physical intuition about the phase diagram and stability of the different phases.  

\subsection{Competition between metal and valley polarized states}
First, we focus on the competition between the metallic state and valley polarized state (setting $\Delta_0 = 0$). The anisotropic part of the interaction $u_1(\q)$, while crucial for the excitonic order parameter, does not play a prominent role here other than altering phase boundaries slightly, so we set it to zero for simplicity. In this case, the covariance matrix takes the form
\beq
\langle c^\dagger_{\tau, \k} c_{\tau^\prime \k^\prime} \rangle = \delta_{\tau, \tau^\prime} \delta_{\k, \k^\prime} f^{\tau}_{\k}, \text{ where } f^{\tau}_{\k} = \Theta(\varepsilon^\tau_F - \varepsilon_{\k})
\eeq
with a separate Fermi energy $\varepsilon^\tau_F = \mu + \tau h$ for the two bands ($\tau = \pm 1$). Therefore, the regularized HF energy density is given by 
\beq
e^{HF}(P_v,0)  = - \frac{1}{2} \sum_{\tau} \int_{\q,\k} u_0(\q) f^\tau_{\k + \q/2} f^\tau_{\k - \q/2} + \sum_{\tau} \int_{\k} \varepsilon_\k f^{\tau}_{\k}
\label{eq:CompFM}
\eeq
To intuitively understand the physics, let us consider two extreme limits. For the fully valley polarized state, one of bands is completely full while the other is completely empty. Hence, Eq.~(\ref{eq:CompFM}) evaluates to
\beq
e^{HF}(1,0) = - \frac{1}{2} \int u_0(\q) + \int_{\k} \varepsilon_\k
\label{eq:eFP}
\eeq
For the unpolarized metal, $f^{+}_\k = f^{-}_\k \equiv f_\k = \Theta(-\varepsilon_\k)$. Defining $g(\q) = \int_{\k} f_{\k + \q/2} f_{\k - \q/2}$, Eq.~(\ref{eq:CompFM}) evaluates to
\beq
e^{HF}(0,0) = - \frac{1}{2} \int u_0(\q) g(\q) + 2 \int_{\k: \varepsilon_\k < 0} \varepsilon_\k 
\label{eq:eUM}
\eeq
The function $g(\q)$ is proportional to the overlap of the Fermi surface with itself when shifted by $\q$. Hence, $g(\q)$ has a maximum value of 1 at $\q = 0$ and decreases with $\q$ till $\q$ is half a reciprocal lattice vector. Since $u_0(\q)$ contains the Landau level projection factor $F^2(\q) = e^{- q^2 l_B^2/2}$, the main contribution to the interaction term comes from $g(\q)$ close to zero, which implies that the unpolarized metal has higher energy than the valley polarized state. In other words, interaction favors valley polarization. On the other hand, the kinetic term from the periodic potential favors the metal, as a full dispersing band costs more energy than two half-filled bands. 

\subsection{Stability of valley polarized insulator and unpolarized metal to exciton vortex lattice}
Following our previous discussion about the metallic phase and the valley polarized insulator, we need to establish that both these phases are stable to an excitonic phase with non-zero $\Delta(\r)$ in presence of anisotropy in the interactions ($u_1(\q) \neq 0$). For two $C = +1$ bands, it is well-known that an infinitesimal anisotropy will drive exciton condensation with uniform magnitude. As argued in the main text, our excitonic order parameter $\Delta(\r)$ formed which has electrons from the $C = +1$ band and holes from the $C = -1$ band, will behave like a superconducting order parameter in presence of a uniform magnetic field. Therefore, we can rule out a uniform exciton condensate, but an exciton vortex lattice indeed remains a distinct possibility. Below, we argue that such a phase is also energetically more expensive as long as the anisotropy is small enough. 

We start off with the fully valley polarized state, corresponding to a large $h$. We now add a small $\Delta_\k$ to see if we gain energy in presence of an arbitrarily weak anisotropy $u_1$, while keeping the filling fixed to half. In this limit, the lower ($\beta$) band is still full while the upper ($\alpha$) band is empty, so we can write $\Theta_{\k,\alpha} = 0$ and $\Theta_{\k,\beta} = 1$. We can write the covariance matrix from Eq.~(\ref{eq:Pmat}) as follows:
 \beq
 P_{\tau, \tau^\prime}(\k) = \frac{1}{2}\begin{pmatrix}
 1 - \frac{h}{\sqrt{h^2 + |\Delta_\k|^2}} & - \frac{\Delta_\k}{\sqrt{h^2 + |\Delta_\k|^2}}  \\
- \frac{\Delta_\k^*}{\sqrt{h^2 + |\Delta_\k|^2}} & 1 + \frac{h}{\sqrt{h^2 + |\Delta_\k|^2}}
 \end{pmatrix}
 \label{eq:P}
 \eeq
Using the form of $P$ from Eq.~(\ref{eq:P}) and writing out the terms in Eq.~(\ref{eq:EHF}) in terms of $u_0$ and $u_1$, we find that (using $\Tr(P(\k)) = 1$):
\beq
e^{HF}(1,\Delta_0) &=& \int_{\k} \varepsilon_\k +  \frac{1}{2} \left( \int_\k \frac{h}{\sqrt{h^2 + |\Delta_\k|^2}} \right)^2 u_1(\mathbf{0}) \nn
& & - \frac{1}{2} \int_{\q,\k} \frac{[u_0(\q) + u_1(\q)]}{2} \left(1 + \frac{h^2}{\sqrt{(h^2 + |\Delta_{\k+ \q/2}|^2 )(h^2 + |\Delta_{\k- \q/2}|^2})} \right) \nn
& & - \frac{1}{2} \int_{\q,\k} \frac{[u_0(\q) - u_1(\q)]}{4} \frac{(\Delta_{\k- \q/2}\Delta^*_{\k+ \q/2} + \Delta^*_{\k- \q/2} \Delta_{\k+ \q/2})}{\sqrt{(h^2 + |\Delta_{\k+ \q/2}|^2 )(h^2 + |\Delta_{\k- \q/2}|^2})}
\label{eq:EHFlargeh}
\eeq
 To check the stability perturbatively, we expand in powers of $|\Delta_0|/h$ and consider the difference of energy density $e^{HF}(1,\Delta_0)$ and $e^{HF}(1,0)$ for the fully polarized state  upto quadratic order. 
\beq
\label{eq:EnDiff1}
e^{HF}(1,\Delta_0) - e^{HF}(1,0) &=& \frac{1}{2}  \int_{\q,\k}  \frac{u_0(\q)}{4} \frac{|\Delta_{\k+ \q/2} - \Delta_{\k- \q/2}|^2}{h^2} + \frac{1}{2}  \int_{\q,\k} \frac{u_1(\q)}{4} \frac{|\Delta_{\k+ \q/2} + \Delta_{\k- \q/2}|^2}{h^2} - \frac{u_1(\mathbf{0})}{2} \int_{\k} \frac{|\Delta_\k|^2}{h^2} \nn
\eeq
The first two terms raise the energy, while the last term lowers the energy of our variational state with respect to the fully valley polarized state. If both bands had the same Chern number, a spatially uniform $\Delta$ is allowed so that $\Delta_\k = \Delta_0 ~\forall ~\k $. In this case, the first term in Eq.~(\ref{eq:EnDiff1}) does not contribute, and we have
\beq
e^{HF}(1,\Delta_0) - e^{HF}(1,0) &=& \frac{|\Delta_0|^2}{2} \int_{\q} \left( u_1(\q) - u_1(\mathbf{0}) \right) 
\eeq
 For a uniform exciton condensate with a spatially uniform $\Delta$, we have $\Delta_\k = \Delta_0 ~ \forall ~ \k$. Therefore, 
\beq
e^{HF}(1,\Delta_0) - e^{HF}(1,0) = \frac{|\Delta_0|^2}{2 h^2} \int_{\q} \left( u_1(\q) - u_1(\mathbf{0}) \right) \nn  
\eeq
Since the Landau level projection adds a factor of $e^{- \q^2 l_B^2/2}$ to the bare anisotropy,  we have $u_1(\q) < u_1(\mathbf{0}) ~ \forall  ~ \q \neq \mathbf{0}$. This negative difference in energy density precisely corresponds to the instability of the fully valley polarized phase of the conventional QHFM to uniform intervalley coherence when $u_1(\q) > 0$. However, for \textit{any} vortex lattice structure, necessitated by topological constraints of hybridizing opposite Chern bands, we have $\Delta_\k$ which is a function of $\k$, Therefore, when $u_1$ is sufficiently small compared to $u_0$ the vortex lattice state has a higher energy than the parent insulator, regardless of the exact nature of the microscopic interactions (as long as both are repulsive). This implies that the fully valley polarized state is robust to the vortex lattice phase. 

\begin{figure}[!t]
 \includegraphics[scale=0.5]{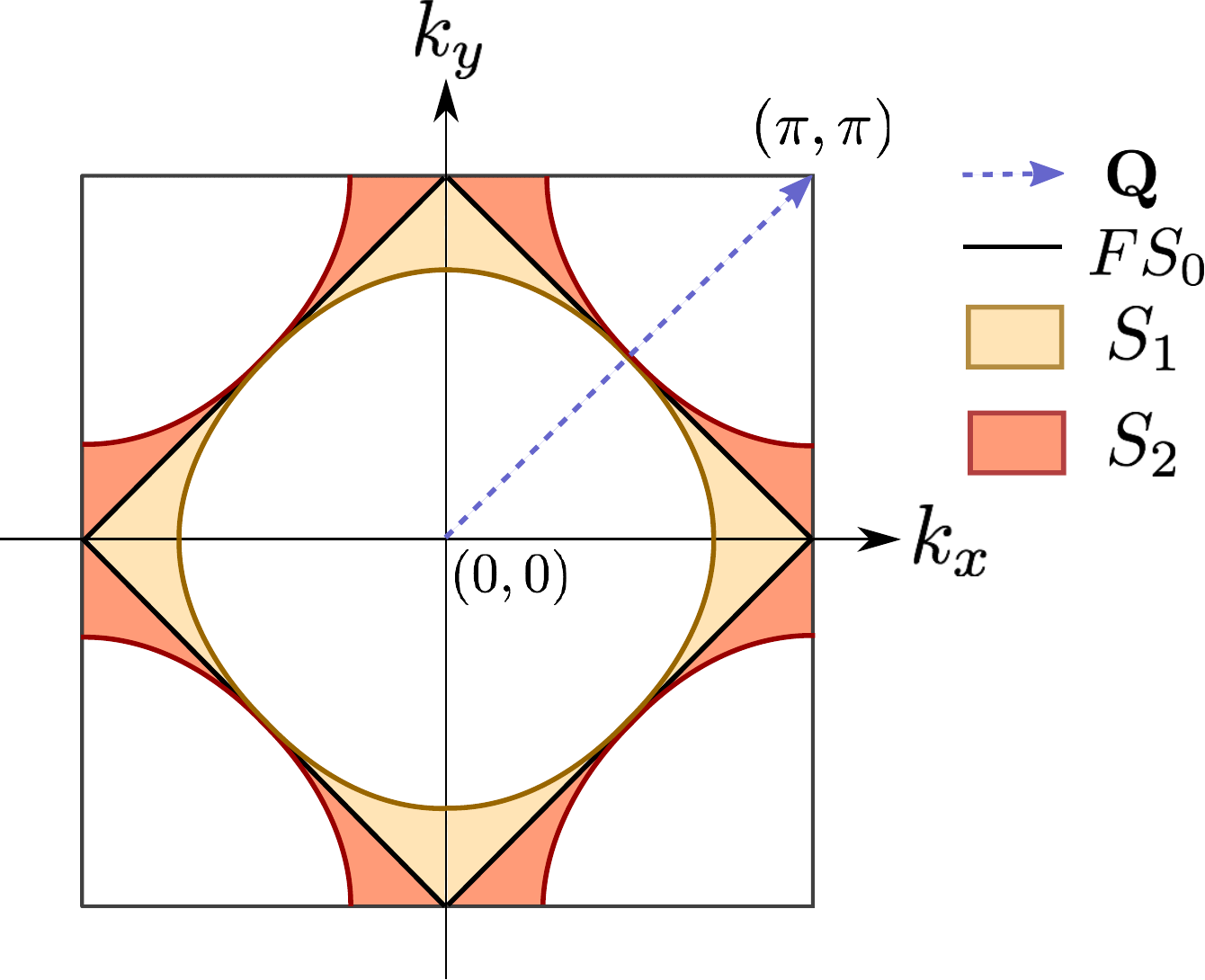}
 \caption{Fermi surfaces of the $\alpha$ and $\beta$ bands when $\Delta_\k \neq 0$, and the regions $S_i$ which correspond to their differences with the original Fermi surface $FS_0$.}
 \label{fig:FS}
 \end{figure}

We now carry out the previous analysis for the unpolarized metal, the second state of our interest. In this case, $P_v = 0$ (obtained by setting $h=0$), so the covariance matrix is given by
\beq
P_{\tau \tau^\prime}(\k) = \frac{1}{2} \begin{pmatrix}
\Theta_{\k,\alpha} + \Theta_{\k,\beta} & e^{i \phi_\k} (\Theta_{\k,\alpha} - \Theta_{\k,\beta}) \\
 e^{-i \phi_\k} (\Theta_{\k,\alpha} - \Theta_{\k,\beta}) & \Theta_{\k,\alpha} + \Theta_{\k,\beta} 
\end{pmatrix}
\eeq
In this case, the HF energy density evaluates to
\beq
e^{HF}(0,\Delta_0) &=& \int_{\k} \varepsilon_\k (\Theta_{\alpha,\k} + \Theta_{\beta,\k}) - \frac{1}{2} \int_{\q,\k} \frac{u_0(\q) + u_1(\q)}{2}  \left( \Theta_{\alpha,-} + \Theta_{\beta,-}\right) \left( \Theta_{\alpha,+} + \Theta_{\beta,+}\right) \nn
 &  & = \frac{1}{2} \int_{\q,\k} \frac{u_0(\q) - u_1(\q)}{2} \cos\left( \phi_{+} - \phi_{-}\right)  \left( \Theta_{\alpha,-} - \Theta_{\beta,-}\right) \left( \Theta_{\alpha,+} - \Theta_{\beta,+}\right) 
 \label{eq:HFMvsEVL}
\eeq
 where the labels $\pm$ are shorthand for momenta $\k \pm \q/2$. We observe that at a total filling of $\nu = 1$, the Fermi surface of the $\beta$ band is identical to the Fermi surface of the $\alpha$ band shifted by $\Q = (\pi, \pi)$. To prove this, we use $\varepsilon_{\k + \Q} = - \varepsilon_\k$ and $\Delta_{\k + \Q} = - \Delta_{\k}$ (as can be seen from the analytical form of $\Delta_\k$ in Eq.~(\ref{eq:VorMom})).
 \beq
 E_{\alpha, \k + \Q} = \varepsilon_{\k + \Q} + |\Delta_{\k + \Q}| = -  \varepsilon_{\k} + |\Delta_\k| = - E_{\beta,\k} \implies \Theta( E_{\alpha, \k + \Q}) = \Theta(-E_{\beta,\k}) = 1 - \Theta(E_{\beta,\k}) 
 \label{eq:rel}
 \eeq
  This has the very important consequence that the chemical potential $\mu$ is fixed to zero at half filling, and the system behaves like a compensated semi-metal with equally sized electron and hole Fermi surfaces. To analyze the energetics, it is convenient to define two subsets of the BZ. Let $FS_0$ be the original diamond shaped Fermi surface of the bands at $\Delta_\k = 0$, defined by the contours $|k_x \pm k_y| =  \pi$ in the first BZ. Then we define (see Fig. \ref{fig:FS})
 \beq
 S_1 = \{\k: ~ \k \in FS_0 \text{ and } E_{\alpha,\k} > 0 \}; ~  S_2 = \{\k: ~   E_{\beta,\k} < 0 \text{ and } \k \notin FS_0 \}
 \eeq 
 
 Note that $\k \in S_1$ implies $\k + \Q \in S_2$, so the area of $S_i$ ($i = 1,2$) in the BZ, which we denote by $\mathcal{A}_{S_i}$ are equal. Using these, we compute the kinectic energy term of the vortex lattice phase. 
  \beq
  \int_{\k} \varepsilon_\k (\Theta_{\alpha,\k} + \Theta_{\beta,\k}) &=& \int_{\k} \varepsilon_\k (\Theta_{\k \in FS_0} - \Theta_{\k \in S_1} + \Theta_{\k \in FS_0} + \Theta_{\k \in S_2}) \nn
    & = &  2 \int_{\k \in FS_0}  \varepsilon_\k  +  2 \int_{\k \in S_2} \varepsilon_\k
 \eeq
 The second term is positive, and denotes the increase of kinetic energy of our variational state by virtue of distorting the bands. The interaction term can also be analogously split into contributions coming from the original Fermi surface, and those coming from Fermi surface distortions. Further, we need to consider the overlap of Fermi surfaces shifted by a momenta of $\q$, but the Landau level projection factors imply that only the overlap at $q \approx 0$ is important when $a \ll l_B$. While our lattice has $a/l_B = \frac{1}{\sqrt{2\pi}}$ so we are not strictly in this limit, it is nevertheless instructive to look at, as the Fermi surface overlaps can be succinctly expressed in terms of $S_1$ and $S_2$.  Adding all contributions, we finally find that the energy density difference is given by:
 \beq
 e^{HF}(0,\Delta_0) - e^{HF}(0,0) =  2 \int_{\k \in S_2} \varepsilon_\k + \frac{ 2 u_1(l_B^{-1}) \mathcal{A}_{S_1}}{l_B^2}
 \eeq
 where we have approximated $u_1(\q)$ by $u_1(l_B^{-1})$ to avoid potential singularities at $q = 0$. For weakly anisotropic repulsive interactions, $u_1 > 0$ so both terms raise the energy of the vortex lattice variational state with respect to the unpolarized metal. For $u_1 < 0$, the first term raises the energy while the second one lowers it. However, in both cases for a small enough $u_1$, the vortex lattice state still has a higher energy and will not be favorable. 
 
 \section{Numerical Hartree-Fock analysis of magic angle graphene}
 
 To confirm that the physical picture discussed in the main text in terms of a Lowest Landau model indeed applies to magic angle graphene, we have numerically solved the Hartree-Fock self-consistency equations in the case where there is a sublattice splitting of $15$ meV on the top graphene layer. We used a dual gate-screened Coulomb potential with a gate distance of $20$ nm, and a dielectric constant of $\epsilon = 9.5$. The twist angle was $\theta = 1.05^\circ$. The simulations were done on a $24\times 24$ momentum grid, keeping six BM bands per spin and valley. The setup of the simulation is exactly the same as that in Ref. \cite{KIVCpaper}, and we refer to that paper for more details. We would like to point out that the only assumption that went into the numerics is that the ground state does not break translation symmetry at the Moire scale -- every other type of symmetry breaking is allowed to occur and no bias is introduced.
 
\begin{figure}
    \centering
    \includegraphics[scale=0.5]{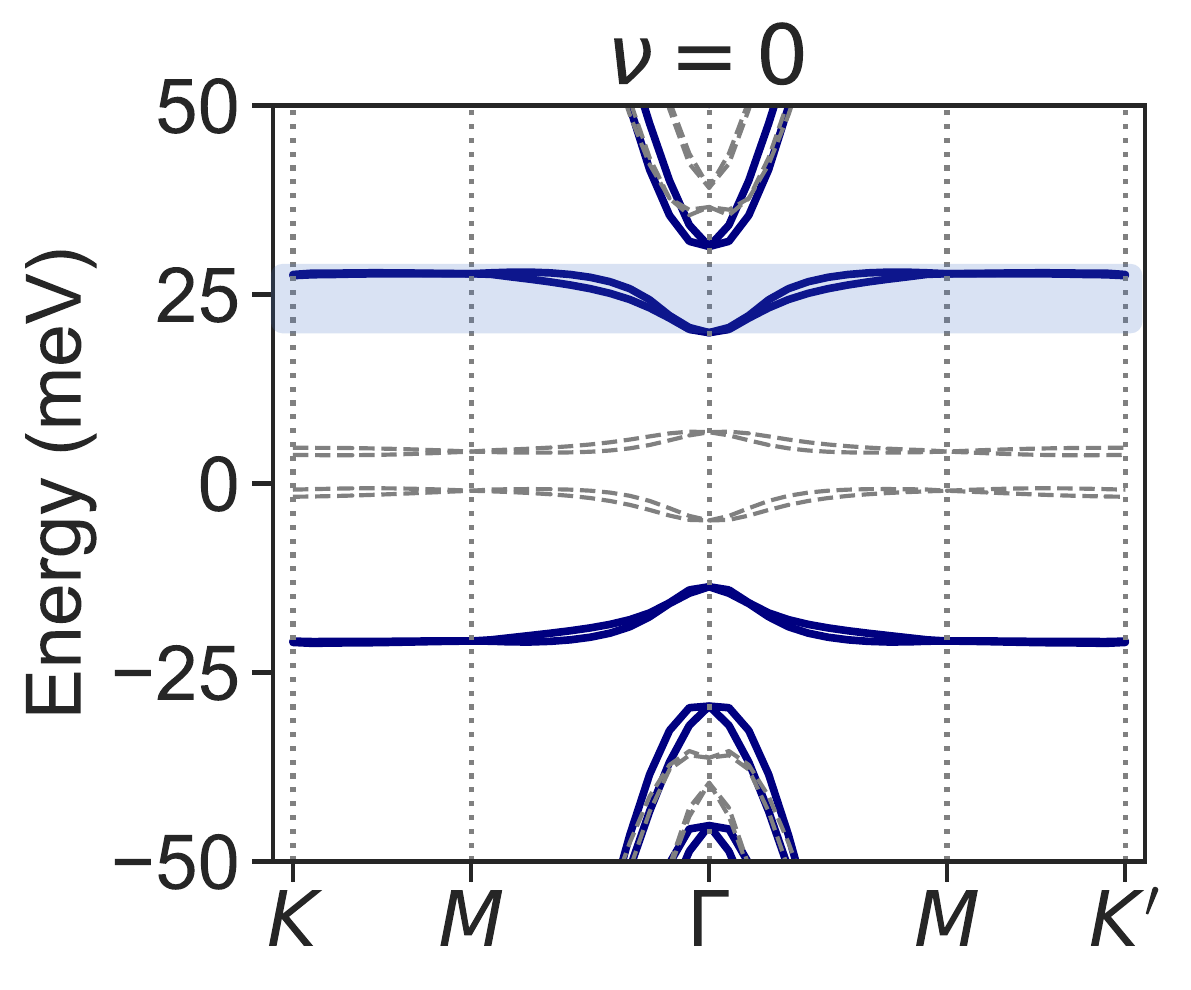}
    \caption{Self-consistent Hartree-Fock band spectrum of twisted bilayer graphene with sublattice splitting $\Delta_t=15$ meV and twist angle $\theta = 1.05^\circ$ at charge neutrality (i.e. $\nu=0$). The active conduction bands, which are approximated by LLL in the main text, are highlighted. The dashed gray lines is the original BM band spectrum.} 
    \label{fig:HFCNP}
\end{figure}

 In Fig. \ref{fig:HFCNP}, we show the self-consistent Hartree-Fock insulating band spectrum at charge neutrality. It has a large bandgap of $\sim 30$ meV. The four lowest conduction bands, which are the bands we focus on in the main text, also remain very flat and do not touch the remote bands. The bandwidth is of the same order as that of the bare BM bands. We also find that at charge neutrality, no symmetries are broken spontaneously. Also, if we initialize the algorithm with the state corresponding to the insulating ground state of the BM model with sublattice splitting, we find that convergence is achieved after only 5 to 10 iterations, which means that this state is already very close to the self-consistent solution. Therefore, the bands do not change significantly in going from the BM ground state to the self-consistent solution. 

\begin{figure}
    \centering
    a) 
    \includegraphics[scale=0.5]{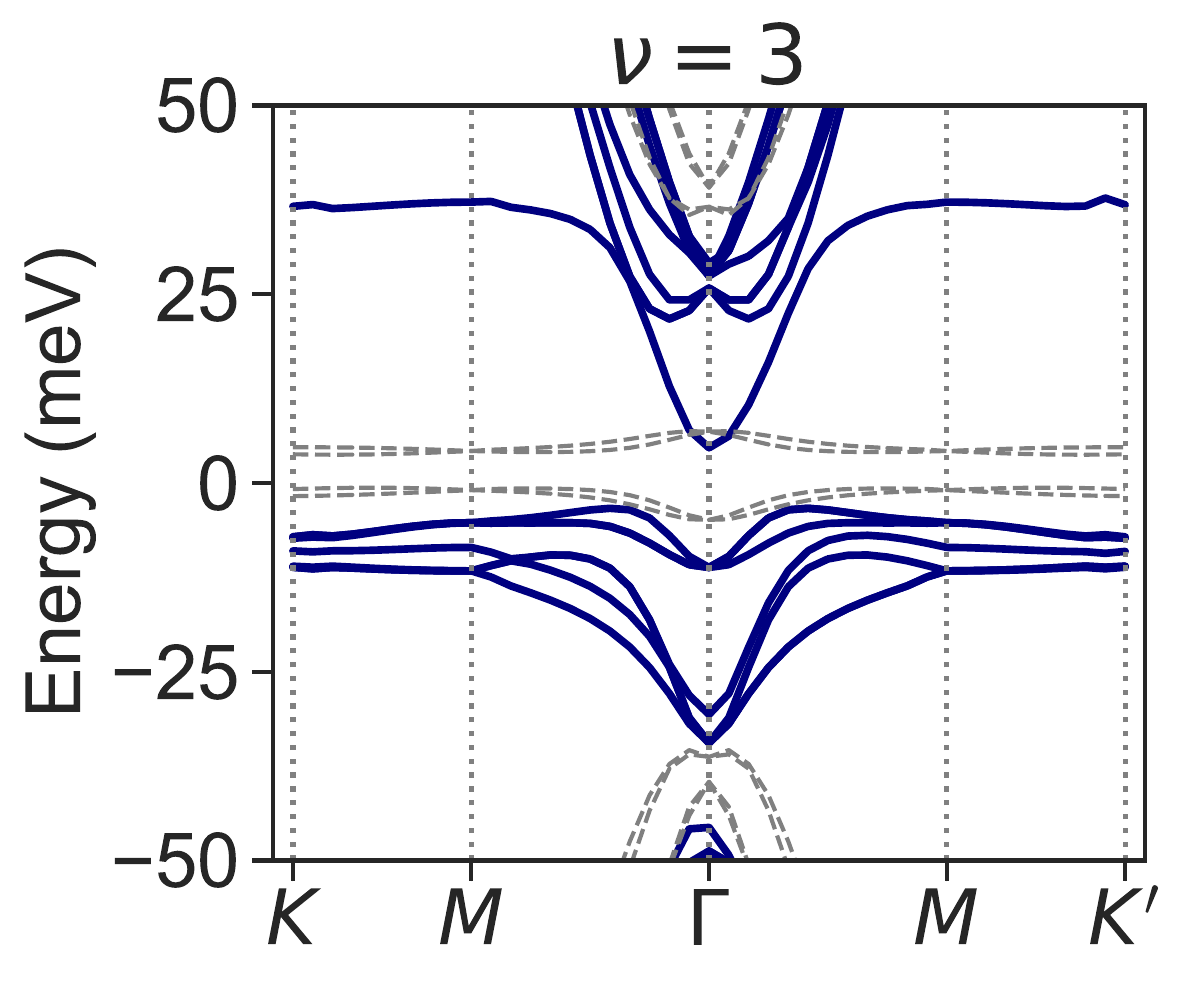}
    \qquad b)
    \includegraphics[scale=0.5]{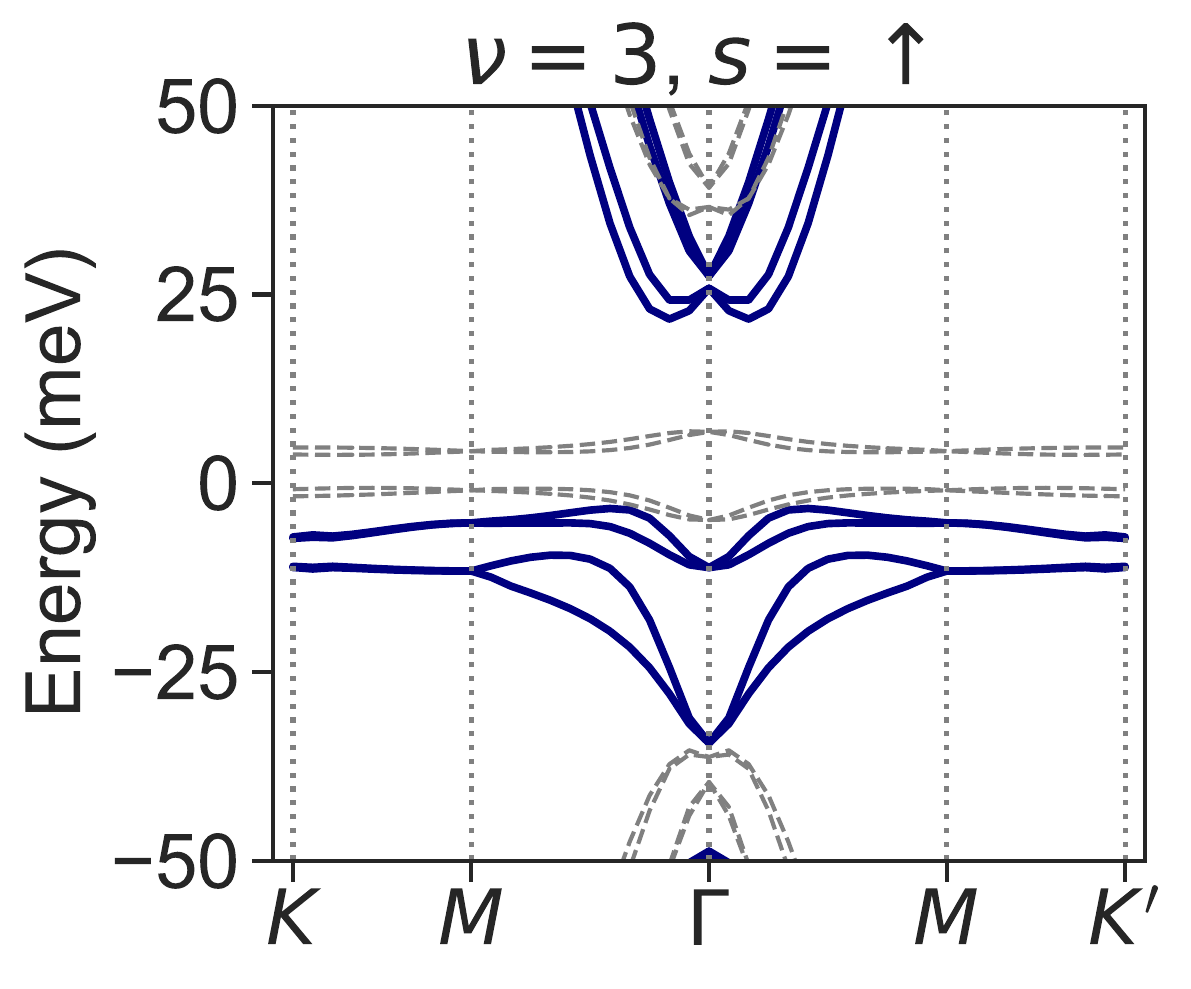}\\
    c)
    \includegraphics[scale=0.5]{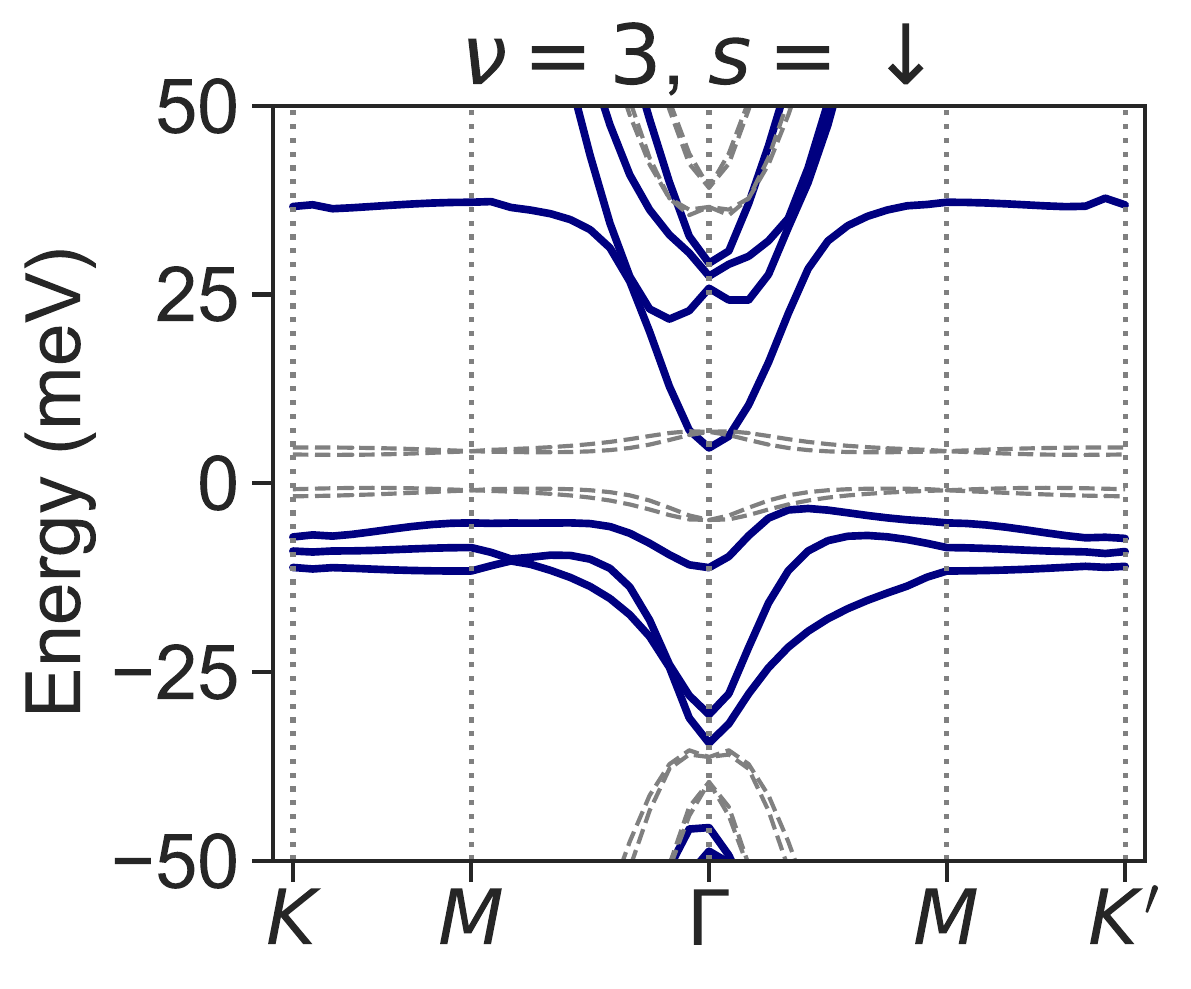}
    \caption{(a) Self-consistent Hartree-Fock band spectrum of twisted bilayer graphene with sublattice splitting $\Delta_t=15$ meV and twist angle $\theta = 1.05^\circ$ at $\nu=3$. The dashed gray lines is the original BM band spectrum. (b) The same as in (a), but only the bands with spin up are shown. (c) Only the bands with spin down are shown.} 
    \label{fig:HFnu3}
\end{figure}

In Fig. \ref{fig:HFnu3}, the self-consistent band Hartree-Fock band spectrum is shown at filling $\nu = 3$. In \ref{fig:HFnu3}(a), the total band spectrum is shown, while in \ref{fig:HFnu3}(b) only the spin up bands are shown, and in \ref{fig:HFnu3}(c) only the spin down bands are shown. Importantly, we find that system has a clear bandgap, and unbroken valley U$(1)$ symmetry. From Fig. \ref{fig:HFnu3} we also clearly see that one spin and valley polarized band of the eight bands around charge neutrality remains unoccupied. This band spectrum therefore agrees with the general physical scenario put forward in the main text.
 
\section{Orbital moment and valley-Zeeman effect}
 
The momentum-dependent orbital moment of electrons in a band labeled by $\alpha$ is given by \cite{XiaoNiu,ChangNiu}
\begin{equation}
m_{\alpha,\textbf{k}}=\frac{e}{\hbar}\sum_{\beta\neq\alpha}\text{Im}\frac{\langle u_{\alpha,\textbf{k}}|\partial_{k_x}H(\textbf{k})|u_{\beta,\textbf{k}}\rangle \langle u_{\beta,\textbf{k}}|\partial_{k_y}H(\textbf{k})|u_{\alpha,\textbf{k}}\rangle}{\varepsilon_{\alpha,\textbf{k}}-\varepsilon_{\beta,\textbf{k}}}\, ,
\end{equation}
where $H(\k)$ is the corresponding Bloch Hamiltonian with eigenstates $|u_{\alpha,\k}\rangle$ and eigenvalues $\varepsilon_{\alpha,\k}$. This orbital moment couples linearly to the out-of-plane component $B^z$ of the magnetic field via the orbital Zeeman term
\begin{equation}
H_{OZ,\alpha} = -\sum_{\k} m_{\alpha,\k} B^z\, .
\end{equation}
The average orbital $g$-factor reported in the main text is given by
\begin{equation}
g_v = \frac{2}{A_{mBZ}\,\mu_B} \int_{\k} m_{+,\k}\, ,
\end{equation}
where $\mu_B$ is the Bohr magneton and $A_{mBZ}$ is the area of the mBZ.

\end{document}